\documentclass[12pt]{iopart}
\usepackage{graphicx}
\usepackage[breaklinks=true,colorlinks=true,linkcolor=blue,urlcolor=blue,citecolor=blue]{hyperref}

\usepackage{color}

\usepackage[utf8x]{inputenc}
\usepackage{subeqn}
\usepackage{wasysym}
\usepackage{tikz}
\usepackage{array}
\usepackage{verbatim}

\usepackage{mathrsfs}
\DeclareSymbolFontAlphabet{\mathrsfs}{rsfs}
\newcommand{\scri}{\mathrsfs{I}}

\newcommand{\pap}{Vano-Vinuales:2014koa}

\newcommand{\CZ}{Z4c}
\newcommand{\alexthesis}{Vano-Vinuales:2015lhj}
\newcommand{\procere}{Vano-Vinuales:2014ada}
\newcommand{\procmg}{Vano-Vinuales:2016mbo}
\newcommand{\papbh}{bhpaper}

\newcommand{\rscri}{r_{\!\!\scri}}
\newcommand{\Kc}{K_{CMC}}

\newcommand{\nconst}{n_{cK}}
\newcommand{\charm}{\xi_{harm}}
\newcommand{\cuplog}{\xi_{1+log}}
\newcommand{\cnK}{\xi_{cK}}
\newcommand{\cpbg}{\xi_\alpha}

\newcommand{\atscrip}[1]{\left.#1\right|_{\scri^+}}
\newcommand{\backbetar}{\hat{\beta^r}}
\newcommand{\backalpha}{\hat{\alpha}}
\newcommand{\backbetarp}{\hat{\beta^{\tilde{r}}}}
\newcommand{\backalphap}{\hat{\tilde{\alpha}}}
\newcommand{\AnilO}{\tilde\Omega}

\begin{document}

\title[Spherically symmetric unconstrained hyperboloidal evolution II: gauge conditions]{Spherical symmetry as a test case for unconstrained hyperboloidal evolution II: gauge conditions}

\author{Alex Vañó-Viñuales$^{1,2}$ and Sascha Husa$^{2}$}
\address{$^1$School of Physics and Astronomy, Cardiff University, Queen's Buildings, The Parade, CF24 3AA, Cardiff, United Kingdom}
\address{$^2$Universitat de les Illes Balears and Institut d'Estudis Espacials de Catalunya, Cra. de Valldemossa km.7.5, 07122 Palma de Mallorca, Spain}
\ead{Alex.Vano-Vinuales@astro.cf.ac.uk}

\begin{abstract}
We present numerical solutions of the hyperboloidal initial value problem  for a self-gravitating scalar field in spherical symmetry, using a variety of standard hyperbolic slicing and shift conditions that we adapt to our hyperboloidal setup. We work in the framework of conformal compactification, and study both evolutions that employ the  preferred conformal gauge, which simplifies the formal singularities of our equations at null infinity, and evolutions without this simplification.
In previous work we have used a staggered grid, which excludes null infinity, while now we include the option of placing a gridpoint directly at null infinity.
We use both the generalized BSSN and conformal Z4 formulations of the Einstein equations, study the effect of different gauge conditions, and show that robust evolutions are possible for a  range of choices.
\end{abstract}



\section{Introduction}

The energy loss and radiation of isolated systems are in general only well defined at future null infinity ($\scri^+$). This is where we can consider  observers of astrophysical events to be located \cite{Barack:1998bv,Leaver1986,PhysRevD.34.384}, and where we would ideally want to extract radiation signals from numerically constructed spacetimes.
In order to treat a physical system of infinite extent, we follow Penrose's approach \cite{PhysRevLett.10.66,Penrose:1965am}, and conformally compactify spacetime: the physical metric $\tilde g_{\mu\nu}$, which diverges at infinity, is rescaled by a conformal factor $\Omega$ that is chosen to vanish at $\scri^+$ to construct a finite metric $\bar g_{\mu\nu}$:
\begin{equation}\label{rescmetric}
\bar g_{\mu\nu} \equiv\Omega^2\tilde g_{\mu\nu} .
\end{equation}
The Einstein field equations written in terms of the rescaled metric $\bar g_{\mu\nu}$ formally diverge at $\scri^+$:
\begin{equation}\label{eq:EEconformal}
G_{\mu\nu}[\bar g] = 8\pi\ T_{\mu\nu} -\frac{2}{\Omega}\left(\bar \nabla_\mu\bar \nabla_\nu\Omega-\bar g_{\mu\nu}\bar \nabla^\gamma\bar \nabla_\gamma\Omega\right)-\frac{3}{\Omega^2}\bar g_{\mu\nu}(\bar \nabla_\gamma\Omega)\bar \nabla^\gamma\Omega \,.
\end{equation}
In our work we will solve an initial value problem where spacetime is foliated by smooth spacelike slices that reach $\scri^+$, known as hyperboloidal slices. The hyperboloidal initial value problem was pioneered by Friedrich, who developed a reformulation of the conformally rescaled Einstein equations \eref{eq:EEconformal} in terms of a regular system of equations, the ``Conformal Field Equations'' \cite{friedrich1983,lrr-2004-1,Friedrich:2003fq}. Due to general relativity's character as a gauge theory, approximation schemes that solve the Einstein equations as an initial value problem tend to suffer from instabilities in the continuum equations. As a consequence, small numerical errors can trigger growing modes in the gauge or constraint degrees of freedom,
see e.g.~\cite{Husa:2002zc,Husa:2005ns,Frauendiener:2004bj}.

In recent years, numerical simulations of mergers of compact objects have provided crucial information about the gravitational wave signals of such events, which was essential to identify the sources the signals discovered since late 2015 by the LIGO and Virgo detectors \cite{Abbott:2016blz,Abbott:2016nmj,TheLIGOScientific:2016wfe,TheLIGOScientific:2016pea}. In these simulations, Cauchy slices are evolved up to some finite distance from the source, and extrapolation methods or characteristic extraction \cite{Boyle:2009vi,Reisswig:2009us} are used to estimate the signal at future null infinity.
Such simulations have only been possible after the breakthroughs of 2005 \cite{Pretorius:2005gq,Campanelli:2005dd,Baker:2005vv}, which have established gauge conditions that are appropriate in particular for the strong field region in the vicinity of black holes, and avoid instabilities that had plagued the field of numerical relativity for many years.
Our goal here is to develop a simple implementation of the hyperboloidal initial value problem: starting with a formulation close to those routinely used to study astrophysical situations like the coalescence of compact objects (such as BSSN and Z4 \cite{PhysRevD.52.5428,Baumgarte:1998te,bona-2003-67}), we then modify it to resolve instabilities rooted in the continuum formulation.
Since the equations are formally singular, numerical schemes to solve the conformally rescaled equations have to be constructed to appropriately regularize the singular terms. A key idea in simplifying such regularization procedures is an appropriate choice of gauge, which is adapted to the inherent simplicity of null infinity. In order to achieve this goal, we face problems and instabilities similar to those that previously plagued the numerics of the strong field region. For a discussion of how such problems are resolved by current gauge conditions on Cauchy evolutions, see e.g. \cite{Hannam:2006vv, Hannam:2008sg}. The problems we face here are mainly due to the compactification of null infinity. As for the problems associated with black holes, the essential problems can already be captured in spherical symmetry, which motivates using spherical symmetry in our present work. While in the full 3-dimensional case new problems may appear, we do expect that resolution of the problems we address in the spherical symmetric context will carry over to the full 3-dimensional case.
The extraction of gravitational wave signals at future null infinity is one of the main final applications of our work. As gravitational waves cannot arise in spherical symmetry, we use a massless scalar field to introduce physical dynamics and mimick gravitational wave effects, so that we can extract and analyse signals in our spherically symmetric implementation. The coupling of a scalar field with the Einstein equations has been widely studied in relation with its critical collapse \cite{Choptuik:1992jv}, using asymptotically flat instead of hyperboloidal slices, where the scalar field does not introduce particular difficulties apart from requiring high resolution to resolve critical phenomena.

A natural class of simple choices for the coordinates and conformal factor at null infinity is known as the ``preferred conformal gauge'' \cite{Tamburino:1966zz,9780511564048,stewart1997advanced}, where the conformal factor $\Omega$ satisfies the wave equation with respect to the unphysical rescaled metric $\bar g_{\mu\nu}$,
\begin{equation}\label{e:prefconfgauge}
\atscrip{\bar\Box\Omega}=0 .
\end{equation}
With this choice one can show that also
\begin{equation}\label{e:ddOmega}
\atscrip{\bar\nabla_\mu\bar\nabla_\nu \Omega}=0 ,
\end{equation}
so that the three divergent terms in \eref{eq:EEconformal} attain a finite limit at $\scri^+$ independently,
and consequently the problem to find a regularized and stable numerical scheme for the hyperboloidal initial value problem simplifies considerably.
The preferred conformal gauge has been used to derive boundary conditions at null infinity for  elliptic-hyperbolic systems, see \cite{Andersson:springer,Rinne:2009qx,Rinne:2013qc},
including constrained evolution schemes in a tetrad formalism \cite{Bardeen:2011ip,Morales:2016rgt}, and alternatively to define gauge conditions for
a hyperbolic free evolution scheme in the approach by Zengino\u{g}lu \cite{Zenginoglu:2007jw,Zenginoglu:2008pw,Zenginoglu:2008wc}.
The preferred conformal gauge ensures that the null generators of $\scri^+$ are geodesic and expansion-free (they are already shear-free in a conformally invariant way). 
A relevant consequence of the expansion-free null generators of $\scri^+$ is that their affine parameter can be identified with the time coordinate at $\scri^+$, which is then commonly called Bondi time and corresponds to the proper time measured by an inertial observer moving along $\scri^+$ \cite{lrr-2004-1}. The Bondi time parameterization can be chosen {\em separately} from the preferred conformal gauge, and such a choice allows to extract signals at $\scri^+$ without any deformations caused by non-inertial coordinate effects.

Pioneering work toward a free evolution scheme for the hyperboloidal initial value problem was performed by
Zengino\u{g}lu \cite{Zenginoglu:2007it}, who carried out evolutions of Schwarzschild initial data in a spherically symmetric code, using the generalized harmonic form of the conformally rescaled Einstein equations (\ref{eq:EEconformal}), and in particular the generalized harmonic coordinate gauge on the conformally compactified manifold. The latter included source functions that satisfied the preferred conformal gauge \eref{e:prefconfgauge}, although it was not clear how to impose the Bondi time parameterization at $\scri^+$. The conformal factor $\Omega$ was a fixed, time-independent function of the coordinates, while the numerical grid did not include $\scri^+$. No numerical instabilities directly associated with null infinity were detected in his work, however continuum instabilities originating in the bulk of the spacetime obstructed long-time evolutions.

Inspired by Zengino\u{g}lu's work, we have previously presented {\em stable} unconstrained evolutions in spherical symmetry for different types of regular initial data \cite{\pap}. Instead of the generalized harmonic formulation of the Einstein equations used by Zengino\u{g}lu, we used the BSSN \cite{PhysRevD.52.5428,Baumgarte:1998te}  and Z4 \cite{bona-2003-67} systems, which have been used successfully for black hole spacetimes in the context of singularity-avoiding slicing conditions \cite{bona}. In this work we did not use the preferred conformal gauge, but
a modified harmonic slicing condition with source terms designed to account for the non-trivial background geometry of hyperboloidal slices (with respect to standard Cauchy slices) and to avoid instabilities in the continuum equations; in particular formally singular damping terms were used to drive the lapse to a constant value at $\scri^+$ throughout the evolution. The shift vector was not evolved, but chosen fixed to its value in flat spacetime and compatible with our requirements at $\scri^+$ (see subsection \ref{ss:basics}).

In the present paper we implement and test a wider class of gauge conditions both for the lapse and the shift. Our lapse conditions are generalisations of the Bona-Mass\'o family of slicing conditions \cite{Bona:1994dr},
while (apart from a fixed shift) we consider modifications of the Gamma-driver shift condition \cite{Alcubierre:2002kk} and the generalized harmonic shift conditions \cite{Friedrich:2000qv}.
As  chosen by Zengino\u{g}lu, we formulate the gauge conditions in terms of the compactified metric instead of the physical one. For work that discusses time-independent hyperboloidal slices using the Bona-Mass\'o family of slicing conditions formulated in terms of the physical (unrescaled) metric, see \cite{Ohme:2009gn}.

We achieve stable evolutions for various combinations of these gauge conditions,
as well as for similar gauge conditions that satisfy the preferred conformal gauge (see \sref{s:pbg}),
and we describe the techniques we have used to suppress instabilities and choose appropriate parameters in our setup to achieve long-term stable evolutions.
While our class of gauge conditions also includes the case of Bondi time parameterization (see sections \ref{ss:prefbon}, \ref{s:pbg} and \ref{ss:regconds}), in this particular case we have not yet been able to suppress code instabilities. We will describe how the solution can in general be reparameterized ``a posteriori'' to obtain undistorted signals at $\scri^+$.
Essentially, we find that while work may be required to make a particular class of gauge conditions work, and in some cases, such as the Bondi time parameterization, further work is still required, a robust treatment of $\scri^+$ is indeed possible with respect to different gauge choices. We will present extensions of our work with black hole spacetimes in future work \cite{\papbh}. For a recent study of hyperboloidal black hole initial data see  \cite{Buchman:2009ew}. Another innovative approach to the hyperboloidal problem that uses the dual foliation formalism \cite{Hilditch:2015qea} is \cite{Hilditch:2016xzh}.

The implementation of stable gauge conditions that satisfy the preferred conformal gauge also simplified the inclusion of $\scri^+$ in the numerical grid. Previously \cite{\pap}, we used a staggered grid that did not include $\scri^+$ (where the evolution equations are formally singular) as a grid point, but we could show that extrapolation to $\scri^+$ converged at the expected order, consistent with the convergence order in the interior of the grid.
With the new non-staggered grid we can evaluate all evolved quantities exactly on future null infinity, without any kind of extrapolation. This approach has in particular two types of technical ingredients (see subsection \ref{ss:nostagimplem}): the identification and imposition of the regularity conditions at $\scri^+$ and the transformation and rewriting of the divergent terms in the equations so that they attain a finite limit at future null infinity.

The paper is organized as follows: a detailed description of the preferred conformal gauge and its relation with the Bondi time is presented in \sref{s:pref}.
How we adapted several standard slicing and shift conditions to the hyperboloidal problem is explained in \sref{s:gau}.
The implementation of the preferred conformal gauge as a system of hyperbolic gauge conditions is described in \sref{s:pbg}.
The performance of some gauge conditions is illustrated in \sref{s:signals}, where the deformation of the scalar field signal at $\scri^+$ is compared for several combinations of slicing and shift conditions.
The implementation of the non-staggered grid that includes a gridpoint on $\scri^+$, as well as some results that compare its behaviour to that of the staggered grid, are presented in \sref{s:ptscri}.
Our notation, details on the formulation of the Einstein equations and the evolution variables we use is summarized in \ref{s:eqs}.
Explicit expressions of the gauge conditions that satisfy the preferred conformal gauge are presented in \ref{a:pbg}, while previous versions of slicing conditions that were tested in our code are included in \ref{a:oldlapse}.

\section{Preferred conformal gauge}\label{s:pref}

Here we will describe our basic coordinate setup (choice of conformal factor and fixed coordinate location of $\scri^+$) as the framework where we consider the preferred conformal gauge \eref{e:prefconfgauge}, and relate both to the Bondi time parameterization at $\scri^+$. How we implemented the preferred conformal gauge as hyperbolic gauge conditions will be explained in \sref{s:pbg}, after introducing several modifications of standard general hyperbolic gauge conditions for the hyperboloidal initial value problem in \sref{s:gau}.

\subsection{Our basic setup: time-independent conformal factor and scri-fixing}\label{ss:basics}

We start by choosing a conformal factor $\Omega$ as a fixed function of the compactified radial coordinate $r$,  so that $\Omega$ is in particular time independent. Furthermore, we assume the following functional form such that $\Omega(r)$ is a regular function which vanishes at $\scri^+$ with non-vanishing derivative (compare e.g. \cite{Husa:2002zc,Schneemann}),
\begin{equation}\label{ein:omega}
\Omega(r) = \left(-\Kc\right)\frac{\rscri^2-r^2}{6\, \rscri},
\end{equation}
where $\rscri$ is the coordinate location of future null infinity (which we set to $\rscri=1$ without restricting generality). The previous expression is obtained in spherical symmetry by imposing an explicitly conformally flat spatial metric in our coordinates (setting (20b) in \cite{\pap}, with $a=-3/\Kc$, equal to unity) on a constant-mean-curvature (CMC) slice, where the trace of the physical extrinsic curvature is constant and corresponds to the negative parameter $\Kc$. 
Note that even if the initial data in a simulation correspond to a CMC slice, this is very likely to change during the evolution, but the value of $\Kc$ in the definition of the conformal factor $\Omega$ \eref{ein:omega} is kept constant throughout the evolution.
An alternative simple choice for a time-independent conformal factor is $\AnilO=\rscri-r$, with $\rscri=1$, used by Zengino\u{g}lu in \cite{Zenginoglu:2006rj,Zenginoglu:2007jw,Zenginoglu:2008wc,Zenginoglu:2008uc}. 
This choice provides a simpler relation between the compactified radial coordinate and the conformal factor ($r=\rscri-\AnilO$, in order to use the latter as a coordinate at $\scri^+$), but it may pose problems at the origin, where the conformal factor is not differentiable. The leading order with which the conformal factor goes to zero at $\scri^+$ is the same in both cases, as our $\Omega\propto(\rscri+r)(\rscri-r)$.

As a further simplification we choose to
set $\scri^+$ to a fixed coordinate location in our numerical grid, which is known as
scri-fixing \cite{Frauendiener:1997ze,Zenginoglu:2007jw}. In our spherically symmetric hyperboloidal setup with a time-independent conformal factor we implement it by making the time coordinate flow along $\scri^+$. This requires that the following relation is satisfied at all times:
\begin{equation}\label{e:scrifix}
\atscrip{-\bar g_{tt}} = \atscrip{\left(\alpha^2-\chi^{-1}\gamma_{rr}{\beta^r}^2\right)}=0,
\end{equation}
see also our discussion and figure 1 in \cite{\pap}.

\subsection{Preferred conformal gauge and Bondi time at $\scri^+$}\label{ss:prefbon}

For a metric and conformal factor that satisfy the preferred conformal gauge condition \eref{e:prefconfgauge} we will use the notation $\hat g_{ab}=\hat\Omega^2 \tilde g_{ab}$,
\begin{equation}\label{e:prefconfgaugehat}
\atscrip{\hat\Box\hat\Omega}=0 .
\end{equation}
A given conformal factor \eref{ein:omega} can always be rescaled to satisfy the preferred conformal gauge condition, by choosing an appropriate $\omega>0$ such that $\hat\Omega=\omega\,\Omega$ and \eref{e:prefconfgaugehat} is satisfied.
From now on we will only refer to \eref{e:prefconfgauge} as the preferred conformal gauge if our setup is such that \eref{ein:omega} satisfies it.
An in-depth description of the preferred conformal gauge can be found in chapter 11 of \cite{Wald} and in \cite{Zenginoglu:2008pw}, and in the same notation as here in section 4.2 in \cite{\alexthesis}.

A real null vector $l^a$ in the Newman-Penrose tetrad \cite{newman1962approach} can be constructed tangent to the null geodesic generators of $\scri^+$, see figure 1 in \cite{\pap}. Written as $\hat l^a = \hat g^{ab}\hat\nabla_b \hat \Omega$, in the preferred conformal gauge it will satisfy the geodesic equation in the affine parameterization $\hat l^a \hat\nabla_a \hat l^b = 0$ on  $\scri^+$ by virtue of
\eref{e:ddOmega}. Its affine parameter is called Bondi time coordinate ($t_B$) at $\scri^+$, scaled as $(\partial/\partial t_B)^a \hat\nabla_a t_B = 1$.
World lines of increasingly distant geodesic observers converge to null geodesic generators of future null
infinity, and proper time converges to Bondi time \cite{lrr-2004-1}, thus the Bondi time parameter can be identified with the proper time of inertial observers at large distances from the source.
Note that the preferred conformal gauge does not imply that the time coordinate is indeed Bondi time at $\scri^+$; this is a separate condition, see the discussion and \eref{extrarel} below, and subsections 2.4 in \cite{Zenginoglu:2007it}.

When the conformal factor \eref{ein:omega} chosen for a numerical construction of the spacetime does not satisfy the preferred conformal gauge, it is still possible to determine a conformal factor that does satisfy \eref{e:prefconfgaugehat} ``a posteriori''.
To do so we take advantage of the conformal freedom introduced by $\omega$ and rewrite \eref{e:prefconfgaugehat} in terms of $\omega$ and our fixed $\Omega$
\begin{equation}
\atscrip{\left[\left(\bar\nabla^a\Omega\right)\left(\bar\nabla_a\ln\omega\right)\right]}=\atscrip{-\frac{\bar\Box\Omega}{4}},
\end{equation}
and solve this equation for $\omega$ using the existing numerical data.
In our spherically symmetric reduction the previous relation takes the form
\begin{equation}\label{e:auxome}
\atscrip{\frac{\dot\omega}{\omega}}=\atscrip{\frac{\beta^r \bar g_{tt}'}{2\alpha^2}} , 
\end{equation}
where a dot denotes a time derivative and a prime a spatial derivative with respect to the compactified radial coordinate. If we substitute \eref{e:scrifix} and use it to rearrange the terms in \eref{e:auxome}, we will obtain (28) in \cite{\pap}.
In the spherically symmetric case we only need to solve a single ordinary differential equation \eref{e:auxome}, but in the 3-dimensional case we would expect a system of equations along the 2-dimensional spherical shell that corresponds to $\scri^+$, which is a much more complicated procedure, and thus evolving with the preferred conformal gauge constitutes an important simplification.

The general relation between our numerical time coordinate $t$, whose scaling is determined by the behaviour of the lapse and shift during evolution, and the Bondi time at $\scri^+$ is given by \cite{\pap}
\begin{equation}\label{e:afinet}
dt_B=\frac{\alpha^2\omega}{\beta^r\Omega'}dt .
\end{equation}
This is obtained by identifying the Bondi time with the inertial time coordinate of flat Minkowski spacetime at $\scri^+$.
If the preferred conformal gauge is satisfied during evolution, the auxiliary conformal factor is trivial, $\omega=1$.
The parameterization of our numerical time coordinate $t$ will depend on the chosen gauge conditions, as they determine the behaviour of $\alpha$ and $\beta^r$ during the evolution. From expression \eref{e:afinet} we obtain the relation between lapse and shift that has to be satisfied at $\scri^+$ for the code time $t$ to coincide with $t_B$ when the preferred conformal gauge \eref{e:prefconfgauge} holds:
\begin{equation}\atscrip{\alpha^2}=\atscrip{\beta^r\Omega'} . \label{extrarel}
\end{equation}
Figure 6 in \cite{\pap} shows a comparison of a signal on some non-affinely parameterized time (and where \eref{e:prefconfgauge} does not hold) and with a Bondi rescaled time. A similar, more complete comparison is included in the present paper in \fref{f:signalsscri}.

As mentioned before, the preferred conformal gauge \eref{e:prefconfgauge} also introduces several simplifications in the equations in our setup, among them:
i) the independent regular limit of the divergent terms at $\scri^+$ in \eref{eq:EEconformal};
ii) the $tt$ component of the rescaled metric goes to zero as $\atscrip{-\bar g_{tt}} \propto\Omega^2$, faster than in the general case where it vanishes as $\atscrip{-\bar g_{tt}}\propto\Omega^1$; 
iii) the compactified radial coordinate $r$ corresponds to the areal radius at $\scri^+$ for all times during the evolution, which in our notation is expressed as $\dot{\bar g_{\theta\theta}}=const$. 

The derivation of hyperbolic gauge conditions that satisfy the preferred conformal gauge \eref{e:prefconfgauge}, is included in \sref{s:pbg}.

\section{Adapting standard hyperbolic gauge conditions to our hyperboloidal approach}\label{s:gau}

The adaptation to the hyperboloidal background geometry is especially important for the equation of motion of the lapse, of which we consider several cases of the Bona-Mass\'o family of slicing conditions, while some additions to the shift evolution equation also improve the simulations. We will explain in detail the slicing and shift conditions used in our approach, and also describe the use of mixed conditions (making some choice of lapse and shift conditions in the interior of the domain and another one near $\scri^+$ - basically to ensure no incoming gauge characteristic speeds appear at $\scri^+$) and how the matching between both was performed.

\subsection{Outline of our approach to construct appropriate gauge conditions}

Here we will follow and extend our approach in \cite{\pap}. Taking advantage of the broad existing knowledge about hyperbolic gauge conditions commonly used in current numerical relativity codes, we adapt them to our hyperboloidal implementation, by adding extra damping terms and source functions, and tuning the introduced parameters experimentally to avoid instabilities.
By ``tune'' we mean the process of finding a range of values for the free parameters (such as the damping strength denoted as $\xi$ in sections \ref{s:gau} and \ref{s:pbg} and in the appendices) that provide a stable numerical evolution of the complete system.

Using a time-independent conformal factor, in particular the simple form we set in \eref{ein:omega}, simplifies the numerical implementation of the hyperboloidal initial value problem significantly: vanishing time derivatives of $\Omega$ simplify the complex and formally singular evolution equations \eref{eq:EEconformal}, and the scri-fixing condition \eref{e:scrifix} prevents the location of $\scri^+$ from moving in the numerical integration domain.

Our approach for adapting existing hyperbolic slicing conditions is the following: we first set the principal part \cite{gustafsson1995time,Calabrese:2005ft,Husa:2007zz}  (the terms in the equations that determine hyperbolicity and the characteristic behaviour of the system), then add damping terms with adjustable strength and source functions, and adapt them according to experimental results until a stable numerical evolution with a well-behaved stationary end state is achieved. The lapse equation of motion can be constructed in the physical spacetime (as was done in \cite{Ohme:2009gn}) and then translated to the conformal domain for its implementation, or be directly developed in the conformal domain.
Both options are essentially equivalent thanks to the simple translation of the quantities between the two domains (see section 2.2.4 in \cite{\alexthesis} and subsection \ref{ss:source} below), but some of the non-principal part terms may have different coefficients; for instance, compare the 4th term in \eref{ae:masterphys}'s and \eref{ae:masterconf}'s right-hand-side (RHS).
As previously Zengino\u{g}lu, we have decided to directly work in the conformal domain, which appeared simpler, but it would be interesting to explore alternatives in future work.
The detailed construction process for the slicing and shift conditions we have tested is presented in the following subsections.

\subsection{Slicing conditions}\label{ss:slicing}

The slicing condition that we use has the same basic form as the one in \cite{\pap}, a generalized Bona-Massó equation
\begin{equation}\label{eg:bonamasso}
\dot \alpha=\beta^r\alpha'-\alpha^2f(\alpha)\left(K-K_0\right)+L_0 .
\end{equation}
In the same way as in \cite{\pap} we add two analytic functions of the radial coordinate to $\dot\alpha$'s RHS: $K_0(r)$ and $L_0(r)$.
For hyperboloidal slices the extrinsic curvature $K$ approaches a finite negative value on future null infinity in our convention ($K$ approaches zero at spacelike infinity for spatially asymptotically flat slices). Setting the usual choice for $f(\alpha)=n\alpha^{(m-2)}$ with $n>0$ in \eref{eg:bonamasso} without $K_0$ yields $\dot \alpha=\beta^r\alpha'-n\alpha^mK+L_0$, and it is clear that for $m=1,2$ the coefficient in front of $\alpha^m$ will be positive, causing an exponential growth that will render the numerical evolution unstable. The quantity $K_0$ is thus included to counteract the effect of the hyperboloidal background geometry.
The source function $L_0$ is calculated from flat data on the hyperboloidal slice and its main purpose is to ensure that the conformal lapse $\alpha$ remains finite at $\scri^+$.
In the following subsections we will describe the three different choices for the function $f(\alpha)$ we have tested: $f(\alpha)=1$, $n_{1+log}/\alpha$ and $\nconst/\alpha^2$.
The slicing condition \eref{eg:bonamasso} can be defined in both the physical or conformal domains, that is, taking $\alpha$ and $K$ to be physical or conformal variables.
In all six cases (three choices of $f(\alpha)$ with the evolution equation set in the physical or conformal domains), the effect of the hyperboloidal geometry is encoded in the source terms (and an extra rescaling by a power of $\Omega$ in \eref{ae:1plogp} and \eref{ae:constp}), so that the hyperbolicity properties of the lapse's evolution equation (assuming reasonable data) are qualitatively the same as those for the Cauchy case - with the exception of $\scri^+$, where the equations are singular.
However, all combinations of the slicing condition considered here form a strongly hyperbolic system of equations in the interior of the domain,
and as $\scri^+$ is an ingoing null hypersurface, no physical information can enter the domain.
In subsection \ref{ss:source} we will describe a possible way of calculating the source functions for a derivation started in the physical domain, as well as giving a similar expression defined in the conformal one.
The precise slicing conditions used in the code, expressed in terms of the actual evolution variables, are also included.
Previous versions of the lapse equation of motion are summarized in \ref{a:oldlapse}.

\subsubsection{Harmonic condition:}\label{s:harmlapse}

The harmonic slicing condition ($f(\alpha)=1$) has been widely studied in the literature and it has already been tested in the hyperboloidal initial value approach in \cite{Zenginoglu:2007it}, together with the generalized harmonic formulation of the Einstein equations. The question of stationary hyperboloidal slices was already raised in \cite{Ohme:2009gn}, where the harmonic condition was considered (without source function $L_0$) and stationary hyperboloidal slices were found. For these reasons we expected that the harmonic lapse condition would provide successful results in our implementation, as was the case dealing with regular spacetimes described in \cite{\pap}. In our formulation, the generalized harmonic slicing condition takes the form
\begin{equation}\label{eg:harmlapse}
\dot \alpha=\beta^r\alpha'-\alpha^2\left(K-K_0\right)+L_0  ,
\end{equation}
and the characteristic gauge speeds provided by this equation of motion are in perfect agreement with the requirements at $\scri^+$, as they coincide with the speed of light. The effect that the tilting of the causal cone along the hyperboloidal slice has on the speed of light is shown in \fref{fr:lightspeeds}, where the characteristic speeds for flat spacetime data are displayed.

\subsubsection{1+log condition:}

The 1+log slicing condition is obtained by substituting $f(\alpha)=\case{n_{1+log}}{\alpha}$ (with $n_{1+log}$ a real positive number) in \eref{eg:bonamasso},
\begin{equation}\label{eg:1ploglapse}
\dot \alpha=\beta^r\alpha'-n_{1+log}\,\alpha\left(K-K_0\right)+L_0 .
\end{equation}
The common choice of $n_{1+log}=2$ has proven to be well-behaved in numerical Cauchy evolutions of spacetimes with strong gravitational fields \cite{Arbona:1999ym,Alcubierre:2001vm,Alcubierre:2002kk}. It has not been directly set in the previous expression because it does not provide physical characteristic speeds at future null infinity for any value of $\Kc$ - a property of the 1+log slicing condition is that gauge speeds can easily become superluminal. If a characteristic speed at $\scri^+$ becomes larger than the lightspeed there, a negative characteristic speed (incoming mode) will also appear at $\scri^+$. The behaviour of the lightspeeds $c_\pm$ in the flat hyperboloidal CMC geometry is illustrated by \fref{fr:lightspeeds}. As $\scri^+$ is an ingoing null hypersurface, no physical information from the outside can cross it. Gauge information is however allowed to enter the domain, but then boundary conditions have to be implemented to treat the incoming gauge modes. Given that boundary conditions are a difficult problem and we do not know what a good choice for them at $\scri^+$ in the hyperboloidal approach would be, we will for the moment use physical gauge propagation speeds.  Thus, instead of $n_{1+log}=2$ we make the following default choice for $n_{1+log}$, which makes the 1+log expression compatible with the physical characteristic speeds at $\scri^+$ in the stationary regime (but not in the dynamical one - see subsection \ref{s:match}):
\begin{equation}\label{eg:nokval}
n_{1+log} = \atscrip{\alpha} = -\frac{\Kc\ \rscri}{3} .
\end{equation}
If \eref{eg:1ploglapse} is defined in the physical domain and translated to conformal evolution quantities, an extra dividing factor $\Omega$ has to added to $n_{1+log}$ to prevent the slicing outgoing characteristic speed from vanishing at $\scri^+$. This was the case for \eref{ae:1plog} in \ref{as:lapseconf}.

The 1+log condition studied in \cite{Ohme:2009gn} did not include a source function $L_0$ as we do here and the calculations were performed with the physical quantities on a non-compactified hyperboloidal slice. Therefore, \cite{Ohme:2009gn}'s conclusions, which state that no stationary hyperboloidal slices could be found for the 1+log case with nonzero offset $K_0$, are not applicable to \eref{eg:1ploglapse} in the way we have constructed it, as the 1+log conditions considered in \cite{Ohme:2009gn} and the current paper are different.
We have indeed performed successful hyperboloidal simulations with our adapted version of 1+log.

\subsubsection{cK condition:}\label{ss:nK}

Motivated by our work to evolve black hole spacetimes, which we will report elsewhere \cite{\papbh} (in particular to increase characteristic speeds inside of the horizon to propagate the perturbations faster, which we found to aid the stability of simulations, see \cite{\procere,\alexthesis}), we also implemented the Bona-Mass\'o condition with $f(\alpha)=\case{\nconst}{\alpha^2}$, where again $\nconst$ is a real positive number,
\begin{equation}\label{eg:nK}
\dot \alpha=\beta^r\alpha'-\nconst\left(K-K_0\right)+L_0 .
\end{equation}
We call this slicing condition the ``cK'' condition, for the form of the $-\nconst K$ term in its RHS: {\it constant times K}. Unlike in the harmonic and 1+log slicing conditions, here the proportionality factor of $K$ in the RHS will not vanish if $\alpha=0$.
In a similar way as with the 1+log condition, the value of $\nconst$ is chosen such that the gauge propagation speeds at $\scri^+$ correspond to the physical ones once the stationary state has been achieved, namely
\begin{equation}\label{eg:nkval}
\nconst = \atscrip{\alpha^2} = \left(-\frac{\Kc\ \rscri}{3}\right)^2 .
\end{equation}
This choice of $f(\alpha)$ does not seem to have been studied in the literature.
A related expression arose in the study of gauge shocks by Alcubierre in \cite{Alcubierre:1996su}, where $f(\alpha)=1+k/\alpha^2$ with $k$ a vanishing or positive constant is the solution of the differential equation that guarantees that the characteristic fields will not generate shocks (as is the case for the harmonic slicing, with $k=0$). It has been further studied in \cite{Alcubierre:2002iq,Reimann:2004wp,Alcubierre:2005gw}, but we are not aware of a studies of the slicing condition \eref{eg:nK} in the literature.
For a vanishing shift, slicing conditions of the form \eref{eg:bonamasso} lead to a generalized wave equation of the lapse along the direction normal to the slices with wave speed $v_g=\alpha\sqrt{f(\alpha)\bar\gamma^{ii}}$ along the direction $x^i$ \cite{Alcubierre:2002iq}, where $\bar\gamma_{ij}$ is the spatial metric. Assuming spherical symmetry (and a flat spatial metric $\bar\gamma^{rr}=\chi/\gamma_{rr}=1$) \cite{Garfinkle:2007yt}, for this choice of $f(\alpha)$ the characteristic speed along the radial direction associated to the slicing condition is $v_g=\sqrt{\nconst}$, a constant. 
In the past, algebraic lapse conditions \cite{1993PhDT.........3B,Anninos:1995am} were considered for black hole spacetimes. Some of those algebraic slicings relate the lapse to the determinant of the spatial metric ($\bar\gamma$ in our notation) and are calculated from the Bona-Mass\'o slicing conditions \eref{eg:bonamasso} assuming a zero shift and $K_0=L_0=0$. Using the relation $\partial_t\log\bar\gamma=-2\alpha K$, for the harmonic case we obtain $\alpha=f(x^i)\sqrt{\bar\gamma}$, where $f(x^i)$ denotes some chosen function of the spatial coordinates, and for the 1+log one, with coefficient $n_{1+log}$ as in \eref{eg:1ploglapse}, $\alpha=f(x^i)+\case{n_{1+log}}{2}\log{\bar\gamma}$. For our cK condition the result is $\alpha=\sqrt{f(x^i)+\nconst\log\bar\gamma}$.

\subsubsection{Calculation of the source functions:}\label{ss:source}

In the physical spacetime we define the following Bona-Mass\'o-like slicing condition, where tildes denote physical quantities:
\begin{eqnarray}\label{ae:masterphysori}
\fl \dot{\tilde\alpha} =& {\beta^{\tilde r}} \partial_{\tilde r}\tilde\alpha -\backbetarp \partial_{\tilde r}\backalphap-\tilde\alpha ^2 f \left(\tilde K -\Kc\right) +\xi_1\backalphap\left(\backalphap-\tilde\alpha\right)+\xi_2\left(\backalphap^2-{\tilde\alpha}^2\right) ,
\end{eqnarray}
The coefficient $f$ is a function of the lapse, the background lapse and the radial coordinate.
The quantities $\backalphap$ and $\backbetarp$ are part of the background physical (uncompactified) metric components (corresponding to the flat metric on a hyperboloidal slice), which together with $\hat{\tilde{\chi}}$, $\hat{{\gamma_{\tilde r \tilde r}}}$ and $\hat{{\gamma_{\theta\theta}}}$ (the last one does not change under a conformal compactification) are given by
{\small
\begin{equation}\label{e:hattildevals}
\fl \hat{\tilde{\chi}} = \hat{{\gamma_{\theta\theta}}} = 1 , \  \hat{{\gamma_{\tilde r\tilde r}}} = \frac{1}{\backalphap^2}, \  \backalphap = \sqrt{1+\left(\frac{\Kc\,\tilde r}{3}\right)^2} \quad \textrm{and} \quad \backbetarp = \frac{\Kc\,\tilde r}{3}\sqrt{1+\left(\frac{\Kc\,\tilde r}{3}\right)^2} .
\end{equation}
}
\noindent The background trace of the extrinsic curvature $\hat{\tilde K}=\Kc$ has already been included in \eref{ae:masterphysori} in the place of $K_0$ in \eref{eg:bonamasso}.
Note that all the previous expressions use the uncompactified radial coordinate $\tilde r$.
The RHS in \eref{ae:masterphysori} has been designed to vanish when the variables $\tilde\alpha$, $\beta^{\tilde r}$ and $\tilde K$ are equal to their background values.
The parameters $\xi_1$ and $\xi_2$ have been introduced to control the damping on the lapse. In most cases, enforcing the linear term controlled by $\xi_1$ is enough for stability. However, including the quadratic damping term with $\xi_2$ is useful when experimenting, as due to the non-linearities the behaviour of the complete system can be different.

The equation of motion for the conformal lapse $\alpha$ is obtained by substituting $\tilde\alpha=\alpha/\Omega$ and $\backalphap=\backalpha/\Omega$. The transformation of the advection terms is such that ${\beta^{\tilde r}} \partial_{\tilde r}\alpha(\tilde r) = {\beta^r} \partial_r\alpha(r)\equiv {\beta^r} \alpha'$. For the relation between physical and conformally compactified quantities see subsection 2.2.4 in \cite{\alexthesis}. We maintain the physical $\tilde K$ (no mixing with the Z4 quantity $\Theta$ is included), but express it as the variation with respect to its background value $\Delta\tilde K = \tilde K-\Kc$. The transformed slicing condition yields
\begin{eqnarray}\label{ae:masterphys}
\fl \dot{\alpha} =& {\beta^r} \alpha '-\backbetar \backalpha'-\frac{\alpha ^2 f \Delta\tilde K}{\Omega }+\left(\frac{\backalpha  \backbetar \Omega'}{\Omega }-\frac{{\alpha} {\beta^r} \Omega '}{\Omega }\right) +\xi_1\backalpha\left(\frac{\backalpha}{\Omega }-\frac{{\alpha}}{\Omega }\right)+\xi_2\left(\frac{\backalpha^2}{\Omega }-\frac{{\alpha}^2}{\Omega }\right) ,
\end{eqnarray}
with $f=1$ for harmonic, $f=m_{1+log}\,\backalpha/\alpha$ for 1+log and $f=m_{cK}\,\backalpha^2/\alpha^2$ for the cK condition - the $\backalpha$ factors are included to make $f$ conformally invariant.
Comparing with \eref{eg:1ploglapse} and \eref{eg:nK}, the coefficients $n_{1+log}\equiv m_{1+log}\backalpha$ and $n_{cK}\equiv m_{cK}\backalpha^2$ are no longer constant: they are time-independent, but functions of the compactified radius $r$.
The components of the background conformally compactified metric are
\begin{equation}\label{e:hatvals}
\hat\chi = \hat{\gamma_{rr}}$ = $\hat{\gamma_{\theta\theta}} = 1 , \quad \hat\alpha = \sqrt{\Omega^2+\left(\frac{\Kc\,r}{3}\right)^2} \quad \textrm{and} \quad \hat\beta^r = \frac{\Kc\,r}{3} ,
\end{equation}
corresponding to initial and stationary flat data and calculated from (20) in \cite{\pap} with $a=-3/\Kc$ and setting $\Omega'$ from \eref{ein:omega}.
The damping terms controlled by $\xi_1$ and $\xi_2$ appear now divided by $\Omega$, in the way that our numerical experiments so far seem to require.

A slicing condition similar to \eref{ae:masterphys} can be derived from \eref{eg:bonamasso} in the conformal domain.
For this, the conformal extrinsic curvature $K$ in \eref{eg:bonamasso} is transformed to its physical equivalent $\tilde K$ using $K = \case{\tilde K}{\Omega}+\case{3\beta^r\Omega'}{\alpha\Omega}$, and then the result is expressed in terms of $\Delta\tilde K$. After adding source functions and damping terms as in \eref{ae:masterphys}, the final expression yields
\begin{eqnarray}\label{ae:masterconf}
\fl \dot{\alpha} =& {\beta^r} \alpha '-\backbetar \backalpha'-\frac{\alpha ^2 f \Delta\tilde K}{\Omega }+3f\left(\frac{\backalpha  \backbetar \Omega'}{\Omega }-\frac{{\alpha} {\beta^r} \Omega '}{\Omega }\right) +\xi_1\backalpha\left(\frac{\backalpha}{\Omega }-\frac{{\alpha}}{\Omega }\right)+\xi_2\left(\frac{\backalpha^2}{\Omega }-\frac{{\alpha}^2}{\Omega }\right) .
\end{eqnarray}
The only difference with respect to \eref{ae:masterphys} is the $3f$ coefficient in front of the 4th term in the RHS above.
If $f=m_{cK}\,\backalpha^2/\alpha^2$ is set and at some point $\alpha=0$, which can happen for strong field data, the slicing condition will have a diverging term and \eref{ae:masterconf} may render the simulation unstable.
This can be solved by introducing the $f$ factor in front of $\alpha^2 \tilde K$ \emph{after} the transformation to the physical trace of the extrinsic curvature has been performed: then the coefficient in front of the 4th term in \eref{ae:masterconf}'s RHS will be simply 3.

\subsection{Shift conditions}

\subsubsection{Fixed shift:}

For our first simulations on regular spacetimes, presented in \cite{\pap}, we used a time-independent shift equal to its initial (and also stationary) value, expressed in terms of our compactified radial coordinate.
This is the simplest shift choice we could find that is compatible with the scri-fixing condition \eref{e:scrifix}.
Nevertheless, for simulations with richer physical content, such as those including a black hole or large perturbations of a scalar field, an evolved time-dependent shift is a more appropriate choice. Among the possible hyperbolic shift conditions, we tested the following ones.

\subsubsection{Gamma-driver:}

Our first test with a hyperbolic shift condition was based on the Gamma-driver shift \cite{Alcubierre:2002kk}, due to its common use in current numerical relativity simulations and its simple form,
\begin{equation}\label{e:Gdspacelike}
\partial_\perp\beta^a = \frac{3}{4}B^a , \quad \partial_\perp B^a = \lambda \partial_\perp\Gamma^a -\eta B^a \quad \textrm{with} \quad \partial_\perp=\partial_t-\beta^i\partial_i ,
\end{equation}
where the auxiliary variable $B^a$ is coupled to the contracted connection $\Gamma^a$.
Given that the flat spacetime value of the three-dimensional contracted connection $\Gamma^r=\gamma^{ij}\Gamma^r_{ij}=-2/r$ is badly behaved at the origin in spherical coordinates, for numerical purposes it is much more convenient to use instead the related  quantity $\Lambda^a$ (see \cite{Brown:2009dd} or \ref{s:eqs} for more details),
\begin{equation}\label{ee:LambGam}
\Lambda^a=\Gamma^a-\gamma^{bc}\hat\Gamma^a_{bc} \quad (+ 2\gamma^{ab}Z_b \ \textrm{in the Z4 formulation}) ,
\end{equation}
where $\hat\Gamma^a_{bc}$ is a time-independent background connection. Thus, we couple the evolution equation for $B^a$ \eref{e:Gdspacelike} to $\Lambda^a$ instead of $\Gamma^a$.
In a similar way as done with the slicing conditions in subsection \ref{ss:source}, we add a source term to account for the stationary value of the advection term, and a damping term. The resulting generalized Gamma-driver shift condition yields 
\begin{subequations}\label{eg:Gammadriver}
\begin{eqnarray}
\dot{{\beta^r}} =&{\beta^r} {\beta^r}'-\backbetar \backbetar' + \frac{3}{4} \mu  {B^r}+ \xi_{\beta^r} \left(\frac{\backbetar}{\Omega }-\frac{{\beta^r}}{\Omega }\right) , \\
\dot{{B^r}} =&  {\beta^r} {B^r}' + \lambda  \left(\dot{\Lambda^r}-{\beta^r} {\Lambda^r}'\right)-\eta  {B^r} ,
\end{eqnarray}
\end{subequations}
The Gamma-driver shift can also be implemented in its integrated version \cite{vanMeter:2006vi}, which in spherical symmetry takes the form
\begin{eqnarray}\label{eg:integGammadriver}
\dot{{\beta^r}} =& {\beta^r} {\beta^r}'-\backbetar \backbetar'+\lambda  \Lambda^r+\eta  (\backbetar-{\beta^r})+ \xi_{\beta^r} \left(\frac{\backbetar}{\Omega }-\frac{{\beta^r}}{\Omega }\right) .
\end{eqnarray}
The properties of the principal part of these equations are formally the same as for their non-hyperboloidal equivalent, because the added source terms do not change the principal part: versions \eref{eg:Gammadriver} or \eref{eg:integGammadriver} of the Gamma-driver together with the Einstein equations and any hyperbolic slicing condition from subsection \ref{ss:slicing} form a strongly hyperbolic system for regular data. The same analysis as in \cite{Gundlach:2006tw} applies here for the quantities evaluated on a hyperboloidal slice (except at $\scri$, where the equations are singular).
The gauge propagation speeds associated with the shift condition depend on the values of the parameters $\lambda$ and $\mu$.  There will be no negative characteristic speeds (no incoming modes) at $\scri^+$ if the product $\lambda\,\mu\leq\left(\case{\rscri}{3}\Kc\right)^2$ for the Gamma-driver condition with auxiliary variable $B^r$ \eref{eg:Gammadriver} and if $\lambda\leq\case{1}{12}\left(\rscri\Kc\right)^2$ for the integrated version \eref{eg:integGammadriver}. The reason for keeping the parameter $\mu$ separated from the parameter $\lambda$ in the Gamma-driver shift condition \eref{eg:Gammadriver} is that even if the characteristic speeds depend only on $\lambda\,\mu$, different choices of $\lambda$ and $\mu$ may still lead to a different numerical behaviour.
For simulations on spatially asymptotically flat slices, the standard choice for the parameter $\lambda$ in the Gamma-driver shift condition is $\lambda=3/4$ (with $\mu=1$), which implies physical characteristic speeds (i.e. coinciding with the speed of light).
A detailed study on parameter choices that provide strong hyperbolicity in non-hyperboloidal black hole simulations can be found in \cite{Gundlach:2006tw}.
In our hyperboloidal approach, setting $\lambda=3/4\,\alpha^2\chi$ in \eref{eg:integGammadriver} gives exactly the physical characteristic speeds everywhere.
Future research, specially when considering a full three-dimensional system, may show that choosing values for the gauge characteristic speeds different from the physical ones is a more convenient option. There is much flexibility in the three-dimensional case, so that better choices may be found, following e.g. \cite{Hilditch:2013ila}. However, at this stage of development of the spherically symmetric case, the choice of physical characteristic speeds for the gauge modes (at least asymptotically) seems like a reasonable option.

\subsubsection{Doubly generalized harmonic shift:}\label{s:harmshift}
The standard generalized harmonic shift condition \cite{Friedrich:2000qv,Alcubierre:2005gh} is defined as the spatial part of the four-dimensional equation $\bar g^{ab}\bar\Gamma^c_{ab}=F^c$, where $F^c$ is usually chosen to be a function of the coordinates and evolution variables, and such that it does not affect the principal part of the Einstein equations.
The time component gives the usual harmonic slicing condition (with dynamical part like \eref{eg:harmlapse}), while in spherical symmetry the radial component gives an evolution equation for the radial shift.
The spatial three-dimensional $\Lambda^a$ \eref{ee:LambGam} (as $\Lambda^r$) is introduced in the shift condition to substitute the first spatial derivatives of the spatial metric components, common practice \cite{Alcubierre:2005gh} required in order to make the system strongly hyperbolic. 
We set the source term to $F^r = \case{\xi_{\beta^r}}{\alpha^2\Omega}\beta^r-\case{L_0}{\alpha^2}-\beta^rF^t$, so that it provides a source function $L_0$ (in a similar way as done with the slicing condition) and a damping term (proportional to $-\beta^r/\Omega$) to ensure that the value of $\beta^r$ stays fixed at $\scri^+$, a behaviour similar to the time-independent shift we used in \cite{\pap}.
Our doubly generalized harmonic shift condition then takes the form:
\begin{equation}\label{eg:harmshift}
\dot \beta^r = \beta^r{\beta^r}'+\alpha^2\chi\Lambda ^r+\frac{\alpha ^2 \chi'}{2 \gamma_{rr}}-\frac{\alpha\chi\alpha'}{\gamma_{rr}}-\frac{2 \alpha^2\chi}{r\gamma_{\theta\theta}} + L_0 -\frac{\xi_{\beta^r}}{\Omega}\beta^r .
\end{equation}
This equation includes a term with $r$ in its denominator that formally diverges at the origin. In numerical tests we find that indeed this term results in an instability at the center $r=0$.
Note also that this term would not have appeared if \eref{eg:harmshift} had been derived from $\bar g^{ab}\bar\Lambda^c_{ab}=F^c$ instead of $\bar g^{ab}\bar\Gamma^c_{ab}=F^c$.
As this diverging term does not belong to the principal part,
we simply absorb it into $L_0$, and thus drop it in equation (\ref{eg:harmshift}).
The final form of our adapted harmonic shift condition is obtained after substituting  $L_0=\backalpha\, \backalpha'-\backbetar \backbetar'+\xi_{\beta^r}\case{\backbetar}{\Omega}$, which ensures that flat hyperboloidal data are a stationary solution of the equation:
\begin{equation}\label{aeg:harmshift}
\dot \beta^r = \beta^r{\beta^r}'-\backbetar \backbetar'+\alpha^2\chi\Lambda ^r+\frac{\alpha ^2 \chi'}{2 \gamma_{rr}}-\frac{\alpha\chi\alpha'}{\gamma_{rr}} +\backalpha\, \backalpha' + \xi_{\beta^r} \left(\frac{\backbetar}{\Omega }-\frac{{\beta^r}}{\Omega }\right)  .
\end{equation}
Note that this evolution equation provides physical characteristic speeds.
Optionally, a damping term of the form $+\eta(\backbetar-{\beta^r})$, like the one in the Gamma-driver condition, can be added.
This damping term has been useful preventing instabilities at the origin for several configurations described in \sref{s:signals}.

\subsection{Matching}\label{s:match}

The numerical results we obtained using the integrated Gamma-driver with $\lambda\neq3/4\,\alpha^2\chi$, the 1+log \eref{eg:1ploglapse} or the cK \eref{eg:nK} slicing conditions, with choices \eref{eg:nokval} and \eref{eg:nkval} respectively, showed, at some point during the evolution, some fluctuations arising at $\scri^+$ that propagated inwards.
An appropriate choice of parameters and enough dissipation can keep these incoming fluctuations under control, but the loss in parameter freedom restricts the numerical experiments considerably.

\begin{figure}[htbp!!]
\center
\includegraphics[width=0.7\linewidth]{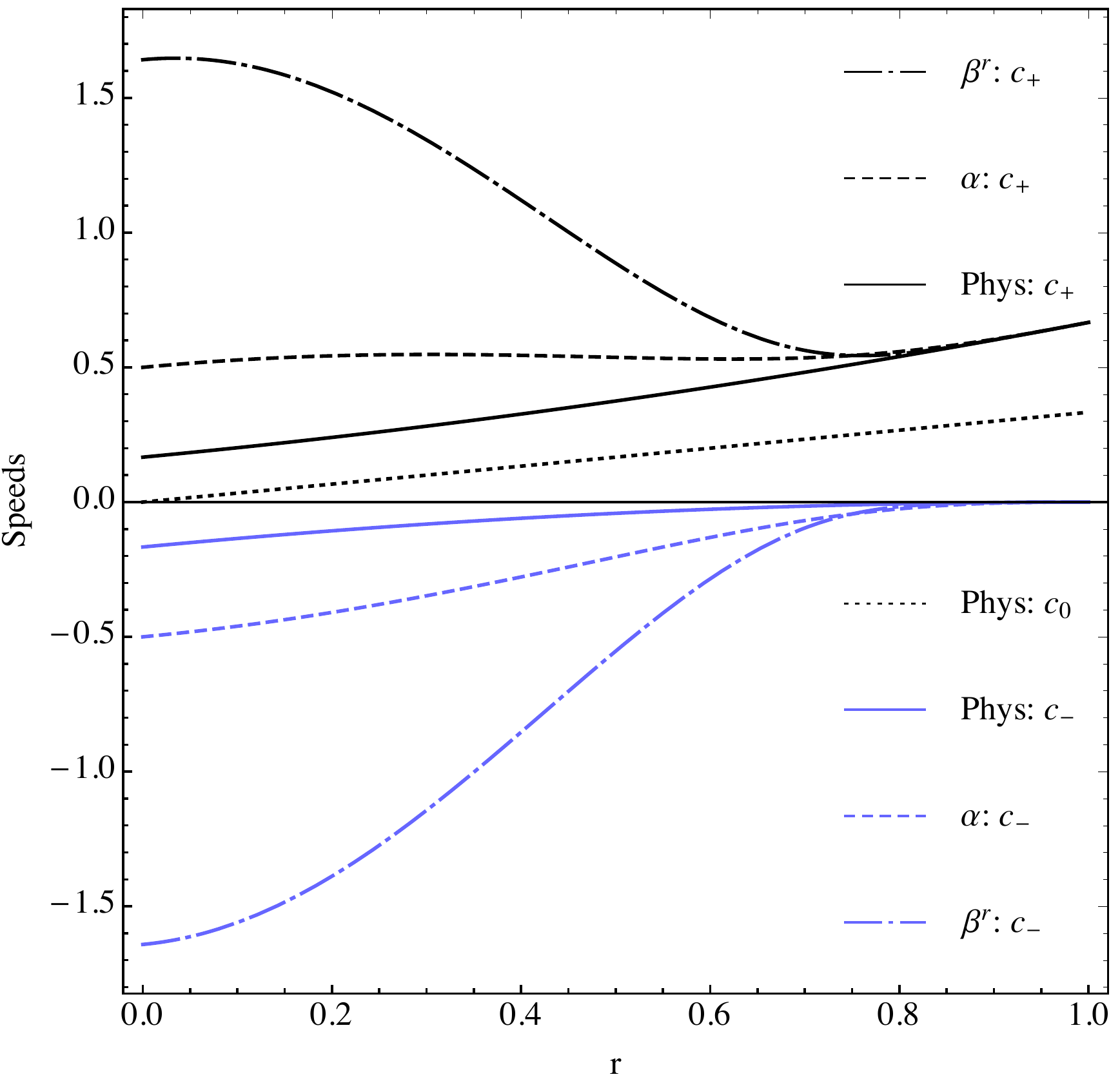}
\caption{Zero speed $c_0=-\beta^r$ and lightspeeds $c_\pm=-\beta^r\pm\alpha\sqrt{\case{\chi}{\gamma_{rr}}}$ plotted for a slice of flat spacetime with $\Kc=-1$ and $\rscri=1$. At $\scri^+$ we have $c_0,c_+>0$ and $c_-=0$. Examples of lapse and shift characteristic speeds with matching \eref{eg:matchf} in \eref{e:superpmatch} ($n_{cK}=2(-\Kc\,\rscri/3)^2$, $n_\alpha$=3, $\lambda=2$ and $n_{\beta^r}=6$) are also included: they coincide with $c_\pm$ at $\scri^+$.}
\label{fr:lightspeeds}
\end{figure}

The most plausible explanation for these fluctuations, as after treating this issue they did not appear anymore, is that at some point at least one of the characteristic speeds at $\scri^+$ becomes negative and some information enters from outside of the domain. This cannot happen to the physical speeds, but it may happen for the gauge speeds: the speeds related to the shift and slicing conditions.
The incoming physical characteristic speed is proportional to the expression in the scri-fixing condition \eref{e:scrifix} (see \fref{fr:lightspeeds}) and this ensures that no information will travel inwards across $\scri^+$.
Nevertheless, the fixed values that are set as coefficients in some principal part terms of the gauge conditions, like \eref{eg:nokval} and \eref{eg:nkval}, and the mentioned ranges for $\lambda$ in \eref{eg:integGammadriver}, may produce an ingoing speed that is no longer proportional to \eref{e:scrifix}, so that non-vanishing incoming gauge speeds can arise when the value of the metric variables at $\scri^+$ differs from the stationary one.
For instance, if $-\left.\sqrt{\frac{\gamma_{rr}}{\chi}}\beta^r\right|_{\scri^+}$ is different from $\sqrt{\nconst}\equiv\alpha|_{\scri^+}$ (given by \eref{eg:nkval}) at some point during the evolution, the speed of the slicing incoming mode at $\scri^+$ does not vanish and uncontrolled information enters the domain.

Considering only physical gauge speeds at $\scri^+$, this problem is solved by making the principal part of the slicing and shift conditions coincide exactly with those of the harmonic slicing and shift conditions respectively at $\scri^+$ at all times during the evolution.
However, we may want to have different characteristic speeds in the interior of the domain, especially when dealing with strong fields near the origin. To achieve this we can use a smooth transition between the required expressions at $\scri^+$ and our choices in the interior.

A possibility is to match using the following function that vanishes at $\scri^+$
\begin{equation}\label{eg:matchf}
f_{match}=\frac{3 \,\rscri}{(-K_{CMC})}\Omega = (\rscri^2-r^2)
\end{equation}
and construct the coefficient of the $K$ or the $\Lambda^r$ term in the corresponding slicing or shift condition respectively as:
\begin{subequations}\label{e:superpmatch}
\begin{eqnarray}
\dot\alpha:& n_{coeff} K \to (n_{cK}{f_{match}}^{n_{\alpha}}+\alpha^2)K , \label{e:supermatchlapse}\\
\dot\beta^r:& \lambda_{coeff} \Lambda^r \to (\lambda{f_{match}}^{n_{\beta^r}}+\alpha^2\chi)\Lambda^r , \label{e:supermatchshift}
\end{eqnarray}
\end{subequations}
where $n_{cK}$, $\lambda$, $n_\alpha$ and $n_{\beta^r}$ are free parameters. In this way, the gauge characteristic speeds at future null infinity will always coincide with the physical speeds and no incoming modes will appear.

The region of matching by superposition in \eref{e:superpmatch} extends from $r=0$ to $r=\rscri$. In order to control the extension and location of the matching region, we have experimented with a new matching that takes place between two given values of the compactified radial coordinate $r_-$ and $r_+$, with $r_-,r_+\in[0,\rscri]$ and $r_-<r_+$. We use a different function,
\begin{equation}\label{eg:matchf2}
f_{match}=\left[1-\left(\frac{r_+^2-r^2}{r_+^2-r_-^2}\right)^4\right]^4 ,
\end{equation}
designed to have a smooth profile everywhere in the domain; assuming the natural extensions from \eref{e:matchinout} ($f_{match}=0$ for $r<r_-$ and $f_{match}=1$ for $r>r_+$), this matching function is $C^3$ at the joining points $r_-$ and $r_+$.
The matching joins $G_{int}$ in the interior part and $G_{ext}$ in the exterior including $\scri^+$,
\begin{equation}\label{e:matchinout}
G = \left\{\begin{array}{lcc} G_{int} & \textrm{if} & r<r_- \\ G_{int} (1-f_{match}) + G_{ext} f_{match} & \textrm{if} & r_-<r<r_+  \\ G_{ext} & \textrm{if} & r_+<r  \end{array}\right\} ,
\end{equation}
where $G$ can be identified with the coefficients $n_{coeff}$ or $\lambda_{coeff}$ in \eref{e:superpmatch} or the complete RHS of $\dot\alpha$ or $\dot\beta^r$.
In this way, for $r<r_-$ we can have larger characteristic speeds, and between $r_+$ and $\rscri$ the physical characteristic propagation speeds will ensure the correct treatment of $\scri^+$.

\section{Implementation of the preferred conformal gauge as hyperbolic gauge conditions}\label{s:pbg}

We will now construct source terms for our generalized harmonic gauge discussed in \sref{s:gau}, which will imply the preferred conformal gauge \eref{e:prefconfgauge}.
The complete derivation and preliminary testing of these gauge conditions is included in subsections 4.5.2 and 7.3.10 in \cite{\alexthesis} respectively.

The main motivation behind the design of these gauge conditions is to try and find a more systematic way of expressing the source functions in the gauge conditions (see \sref{s:gau} for how the source functions were set in the adapted gauge conditions) and also to enforce the preferred conformal gauge \eref{e:prefconfgauge} at all times by means of the gauge conditions, as suggested by \cite{Zenginoglu:2008pw}.
We start with the following expression
\begin{equation}\label{eg:gharmpb}
\tilde\Lambda^c=\tilde F^c , \qquad \textrm{with}\quad \tilde\Lambda^c=\tilde g^{ab}\tilde\Lambda^c_{ab}  \quad \textrm{and} \quad \tilde\Lambda^c_{ab} = \tilde\Gamma^c_{ab}-\hat{\tilde\Gamma}{}^c_{ab} ,
\end{equation}
where on the physical 4-dimensional spacetime we set the contraction of the difference between the connection $\tilde\Gamma^c_{ab}$ of the physical metric $\tilde g_{ab}$ and that $\hat{\tilde\Gamma}{}^c_{ab}$ of a time-independent physical background metric $\hat{\tilde g}_{ab}$ equal to some source function $\tilde F^a$ -- note that these are all 4-dimensional quantities.
This construction basically corresponds to the generalized harmonic gauge condition (see subsections \ref{s:harmlapse} and \ref{s:harmshift}) on the physical spacetime, but with the extra subtraction of background terms (especially convenient in spherical symmetry and on a hyperboloidal background), which was inspired by the definition of the 3-dimensional conformal spatial quantity $\Lambda^a$ (see \ref{s:eqs} and \cite{Brown:2009dd}). The background terms do not change the principal part of the system (the same as for generalized harmonic with respect to the physical metric), because only non-principal part terms including inverse metric components are introduced.
An equivalent construction can be performed in the conformal domain, that is, using the conformal metric $\bar g_{ab}$ and the conformal connections $\bar \Gamma^c_{ab}$ and $\hat{\bar \Gamma}{}^c_{ab}$ in \eref{eg:gharmpb}. However, the obtained numerical results were not satisfactory: stable evolutions were achieved, but at the price of not satisfying the preferred conformal gauge. Thus, instead of including the details here, we point the interested reader to subsections 4.5.1 and 7.3.9 in \cite{\alexthesis}, where the derivation and numerical experiments of the conformal version of \eref{eg:gharmpb} are described in detail.

Using \eref{rescmetric} and the corresponding transformation from physical to conformal Christoffel symbols (see (2.4) in \cite{\alexthesis}, for instance), the above condition is transformed to conformal quantities:
 \begin{equation}\label{eg:gharmpbtranf}
\bar\Lambda^c=-\frac{\partial_d\Omega}{\Omega}\left(4\bar g^{cd}-\bar g^{ab} \hat{\bar g}_{ab} \hat{\bar g}{}^{cd}\right) + \frac{\tilde F^c}{\Omega^2} , \quad \textrm{with}\quad \bar\Lambda^c=\bar g^{ab}\left(\bar\Gamma^c_{ab}-\hat{\bar\Gamma}{}^c_{ab}\right),
\end{equation}
where $\hat{\bar g}_{ab}$ is the time-independent conformal background metric and $\hat{\bar\Gamma}{}^c_{ab}$ its associated connection.
The preferred conformal gauge \eref{e:prefconfgauge} can be expressed as
\begin{equation}\label{eg:prefdecomp}
\bar\Box\Omega = \bar g^{ab}\bar\nabla_a\bar\nabla_b\Omega= \bar g^{ab}\partial_a\partial_b\Omega-\bar g^{ab}\bar\Gamma^c_{ab}\partial_c\Omega ,
\end{equation}
and substituting $\bar\Gamma^c_{ab}$ above from the gauge conditions \eref{eg:gharmpbtranf} yields
\begin{eqnarray}\label{eg:prefdecomprp}
\bar\Box\Omega &=& \bar g^{ab}\partial_a\partial_b\Omega-\bar g^{ab}\hat{\bar\Gamma}{}^c_{ab}\partial_c\Omega +\frac{\partial_c\Omega\partial_d\Omega}{\Omega}\left(4\bar g^{cd}-\bar g^{ab} \hat{\bar g}_{ab} \hat{\bar g}{}^{cd}\right) -\frac{\tilde F^c\partial_c\Omega}{\Omega^2} \nonumber \\
&=&  \bar g^{rr}\Omega''-\bar g^{ab}\hat{\bar\Gamma}^r_{ab}\Omega'+\frac{(\Omega')^2}{\Omega}\left(4\bar g^{rr}-\bar g^{ab} \hat{\bar g}_{ab} \hat{\bar g}{}^{rr}\right) -\frac{\tilde F^r\Omega'}{\Omega^2}.
\end{eqnarray}
The quantity $\atscrip{\Omega'}\neq0$ by definition and in principle also $\atscrip{\Omega''}\neq0$. The metric component $\bar g^{rr}$ in our variables is $\bar g^{rr}=\frac{\chi}{\gamma_{rr}}-\frac{{\beta^r}^2}{\alpha^2}$, which vanishes at $\scri^+$ by virtue of the scri-fixing condition, see \eref{e:scrifix}. The background Christoffel symbol $\hat{\bar\Gamma}{}^r_{ab}$ is calculated from flat spacetime data and it is such that at $\scri^+$ the only nonzero component is $\atscrip{\hat{\bar\Gamma}{}^r_{rr}}=\atscrip{\case{1}{r}}=\case{1}{\rscri}$. Thus, $\bar g^{ab}\hat{\bar\Gamma}{}^r_{ab}\equiv\bar g^{rr}\hat{\bar\Gamma}{}^r_{rr}$ will always vanish at $\scri^+$.
The third term also becomes zero at null infinity, because $\atscrip{\bar g^{rr}}\equiv\atscrip{\hat{\bar g}{}^{rr}}=\mathcal{O}(\Omega^2)$ and $\bar g^{ab} \hat{\bar g}_{ab}$, a combination of conformally rescaled metrics, is expected to attain a finite value. The only remaining term will also vanish at future null infinity if  $\atscrip{\tilde F^r}\propto\Omega^q$ with $q>2$.

The actual evolved gauge conditions for the lapse and shift are obtained by isolating $\dot\alpha$ and $\dot\beta^r$ from the time and radial components of \eref{eg:gharmpbtranf} respectively and are presented in \ref{a:pbg}.
The choice of source function $\tilde F^a$ compatible with the preferred conformal gauge that provided a stable numerical behaviour in our experiments was $\tilde F^r=0$ and $\tilde F^t = \cpbg\Omega(\alpha-\hat\alpha)$, where $\cpbg$ is a parameter to be tuned experimentally.

The relation between lapse and shift at $\scri^+$ for the previously derived gauge conditions is given by \eref{e:regbetar}. A vanishing $\cpbg$ would make the code time $t$ and the Bondi time $t_B$ in \eref{e:afinet} exactly equal, as \eref{e:regbetar} with $\cpbg=0$ reduces to \eref{extrarel}.
This means that we would be able to have the Bondi time parameterization at $\scri^+$ during the numerical evolutions.
Unfortunately, the choice $\cpbg=0$ is not yet possible in our implementation, because it does not provide stable numerical simulations. Only a range of non-negative values is allowed, as $\cpbg$ is introduced as the strength of a damping term that would probably create an exponential growth if the wrong sign was used. The range $\cpbg\in[2,4]$ works fine for most configurations. The conclusion is that despite satisfying the preferred conformal gauge \eref{e:prefconfgauge}, the time coordinate at $\scri^+$ in our simulations will not be affinely parameterized at all times (but it can be easily rescaled using \eref{e:afinet}).

If a match of type \eref{e:superpmatch} is attempted with the gauge conditions derived here and included in \ref{a:pbg}, the exponents $n_\alpha$ and $n_{\beta^r}$ have to be equal to or larger than 3, or else the preferred conformal gauge will not hold at $\scri^+$.

When used together with our spherically symmetric reduction of the Z4c formulation of the Einstein equations, the previously derived gauge conditions provide an evolution that satisfies the preferred conformal gauge at $\scri^+$, in the sense that the scheme converges when increasing the numerical resolution.
However, evolving the same gauge conditions with our implementation of the generalised BSSN formulation does not satisfy \eref{e:prefconfgauge} yet. We are currently working towards obtaining a stable evolution of BSSN together with the preferred conformal gauge.

\section{Gauge condition results}\label{s:signals}

The gauge conditions described in the sections \ref{s:gau} and \ref{s:pbg} provide many possible combinations: slicing and shift conditions that can optionally be matched to others in the region close to $\scri^+$ with a variety of matching functions.
Among the parameters to set we have: the strength of the damping in the slicing and the shift conditions; the coupling in the gauge characteristic speeds for 1+log, cK and Gamma-driver conditions; the location of the possible matchings.
It is beyond the scope of this work to describe all stable configurations with our current spherically symmetric implementation of the Einstein equations (see \ref{s:eqs} for details), for the whole system is interrelated and  changing one parameter can affect the values of many others. We will provide a few examples of gauge choices that appear convenient for simulations of regular data and briefly comment on their observed behaviour.

Our implementation uses the Method of Lines, with 4th order finite differences for the spatial discretization and a 4th order Runge-Kutta for the time integration.
The simulations described here were performed with a staggered grid (right diagram in \fref{grids}), 200 spatial gridpoints and a time step of $dt=0.0005$, which provides a Courant number of 0.1 (we can evolve the staggered grid with a Courant number of 0.2, but the simulations with a point on $\scri^+$ presented in \sref{s:ptscri} require 0.1 for stability). One point off-centered stencils were implemented in the derivatives in the advection terms, as is common practice in numerical relativity simulations \cite{Husa:2007hp}.
Kreiss-Oliger dissipation terms \cite{kreiss1973methods} as used in \cite{Babiuc:2007vr} are added to the RHSs, with a strength parameter of $0.25$ for these simulations.
We evolved our Z4c system (dropping the non-principal part terms labelled with $C_{Z4c}$), with parameters $\kappa_1=1.5$ and $\kappa_2=0$.
The dynamics in the simulations was introduced by a Gaussian-like scalar field of the form $\bar\Phi_0= \tilde\Phi_0/\Omega =A/\Omega\,\e^{-\frac{(r^2-c^2)^2}{4\sigma^4}}$ (like the one presented in figures 5 and 6 in \cite{\pap}) that perturbed flat initial data, with parameter choices $A=0.058$, $\sigma=0.1$ and $c=0.25$, and with the constant trace of the physical extrinsic curvature in the initial data set to $\Kc=-3$. The coordinate location of future null infinity is set to $\rscri=1$.

The gauge conditions that satisfy the preferred conformal gauge \eref{eg:harmtwobpsimpl} provide stable results with the parameter choice $\xi_\alpha=2$.
Both the physical \eref{ae:masterphys} and conformal \eref{ae:masterconf} slicing conditions give good results with $\xi_1=2$ and $\xi_2=0$ or with $\xi_1=0$ and $\xi_2=2$. Choosing $\xi_1=1$ and $\xi_2=0$ together with a fixed shift does not provide enough damping to suppress the small fluctuations that appear in the variable $\Lambda^r$ right next to $\scri^+$ (in a staggered grid the outermost gridpoint is half a spatial step away from $\scri^+$, see \fref{grids}) and propagate inwards while growing in amplitude. Nevertheless, $\xi_1=1$ is enough for long-term stability when both the harmonic lapse and shift are used near $\scri^+$. 

Even in regular spacetimes, instabilities at the origin (in the form of high frequency fluctuations that grow very fast) have appeared for the hyperbolic shift conditions where the coefficient in front of $\Lambda^r$ is proportional to $\alpha^2\chi$, that is, for the harmonic shift and the integrated Gamma-driver with $\lambda=3/4\,\alpha^2\chi$. These instabilities can be effectively suppressed by using a superposition of the form \eref{e:supermatchshift} with $\lambda=0.1$ and $n_{\beta^r}=2$. As a Gamma-driver-type shift condition with constant coefficient $\lambda$ is a better choice for the interior part where $r=0$ or where a black hole would be located, we have not performed an in-depth study of this instability.
More interesting is the performance of the same shift conditions (harmonic shift and integrated Gamma-driver with $\lambda=3/4\,\alpha^2\chi$) at $\scri^+$: the first one is stable at long term and presents smooth features at future null infinity, while the second one requires a damping of at least $\xi_{\beta^r}=2$ (a non-zero value of $\eta$ is not enough) to suppress ingoing fluctuations in $\Lambda^r$ like those described in the previous paragraph. Even if the shift characteristic speeds are the same for the two shift conditions considered, the difference in behaviour can be due to their different characteristic fields or their different coupling to $\Lambda^r$.

The matching of the form \eref{e:superpmatch} can be applied at the level of the coefficient in front of $K$ or $\Lambda^r$, but also to the whole gauge condition, as $G=G_{int}f_{match}^n + G_{ext}$ (using the same notation as in \eref{e:matchinout}). Implementing the latter for the cK condition and the integrated Gamma-driver as $G_{int}$ with $n=3$ and the preferred conformal gauge conditions \eref{eg:harmtwobpsimpl} as $G_{ext}$ provided a stable evolution for the regular spacetime case considered. However, the relaxation to the final stationary state corresponding to flat spacetime was slower ($t_{relax}\sim30$) than when using only the unmatched $G_{int}$ ($t_{relax}\sim6$), for instance.
The other matching considered \eref{e:matchinout} works quite nicely if $r_-\leq0.4$, although the relaxation to the final state is also quite slow ($t_{relax}\sim30$). If $r_->0.4$, the evolution variables do not move towards their final state monotonically, but oscillate around it (this is especially visible in $\Lambda^r$) either with decreasing amplitude, so that the final state is reached in the end, or with an increasing one, which renders the simulation unstable.

\subsection{Deformation of the scalar field signals at $\scri^+$}

\begin{figure}[htbp!!]
\center
\hspace{-5ex}
\includegraphics[width=1.05\linewidth]{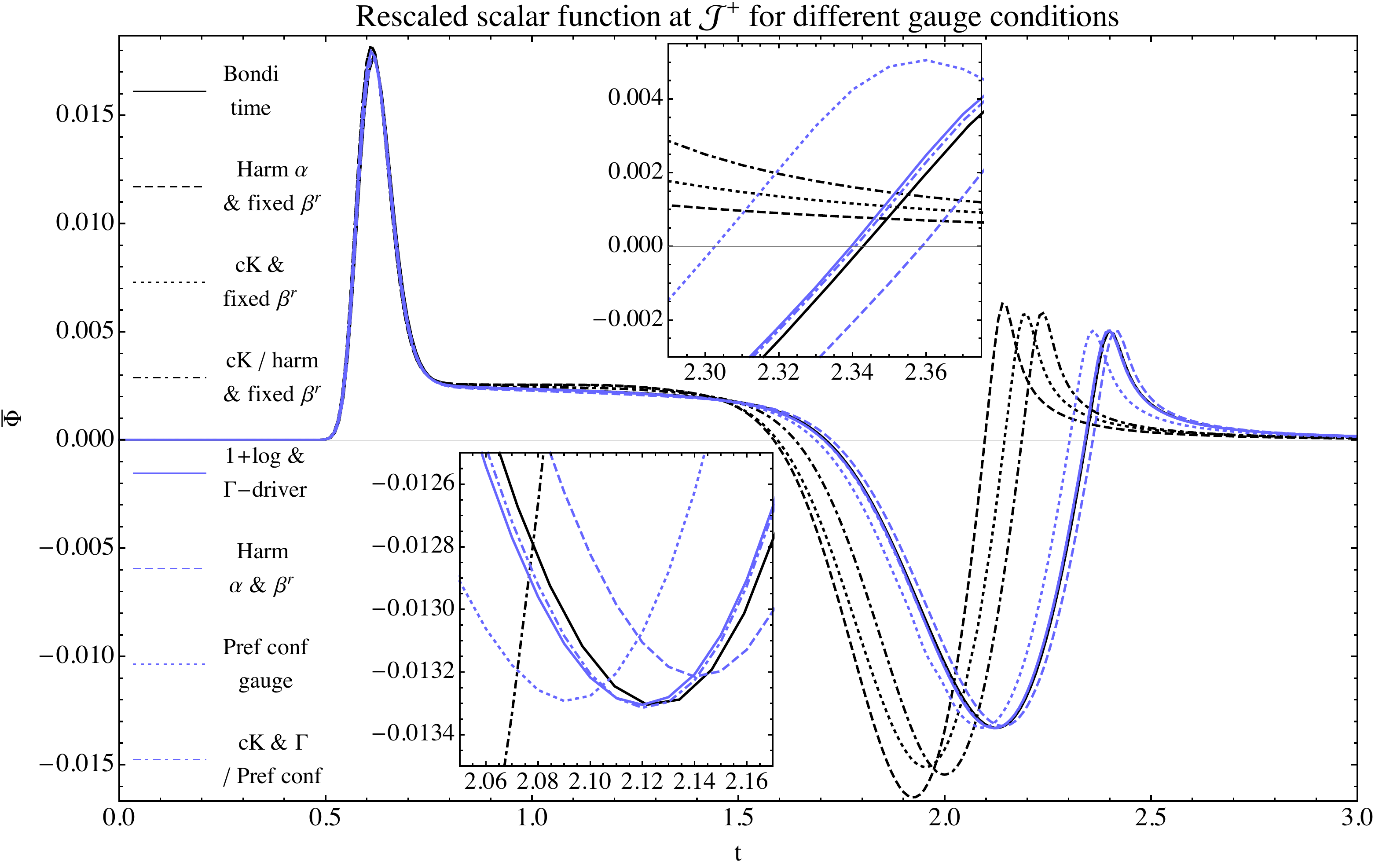}
\vspace{-5ex}
\caption{Signals of the rescaled scalar field $\bar\Phi$ at $\scri^+$ for different combinations of gauge conditions. See the main text for details on each of the gauge choices.}
\label{f:signalsscri}
\end{figure}

\begin{figure}[htbp!!]
\center
\begin{tabular}{@{}c@{}@{}c@{}@{}c@{}}
  \hspace{-3ex}
	\includegraphics[width=0.34\linewidth]{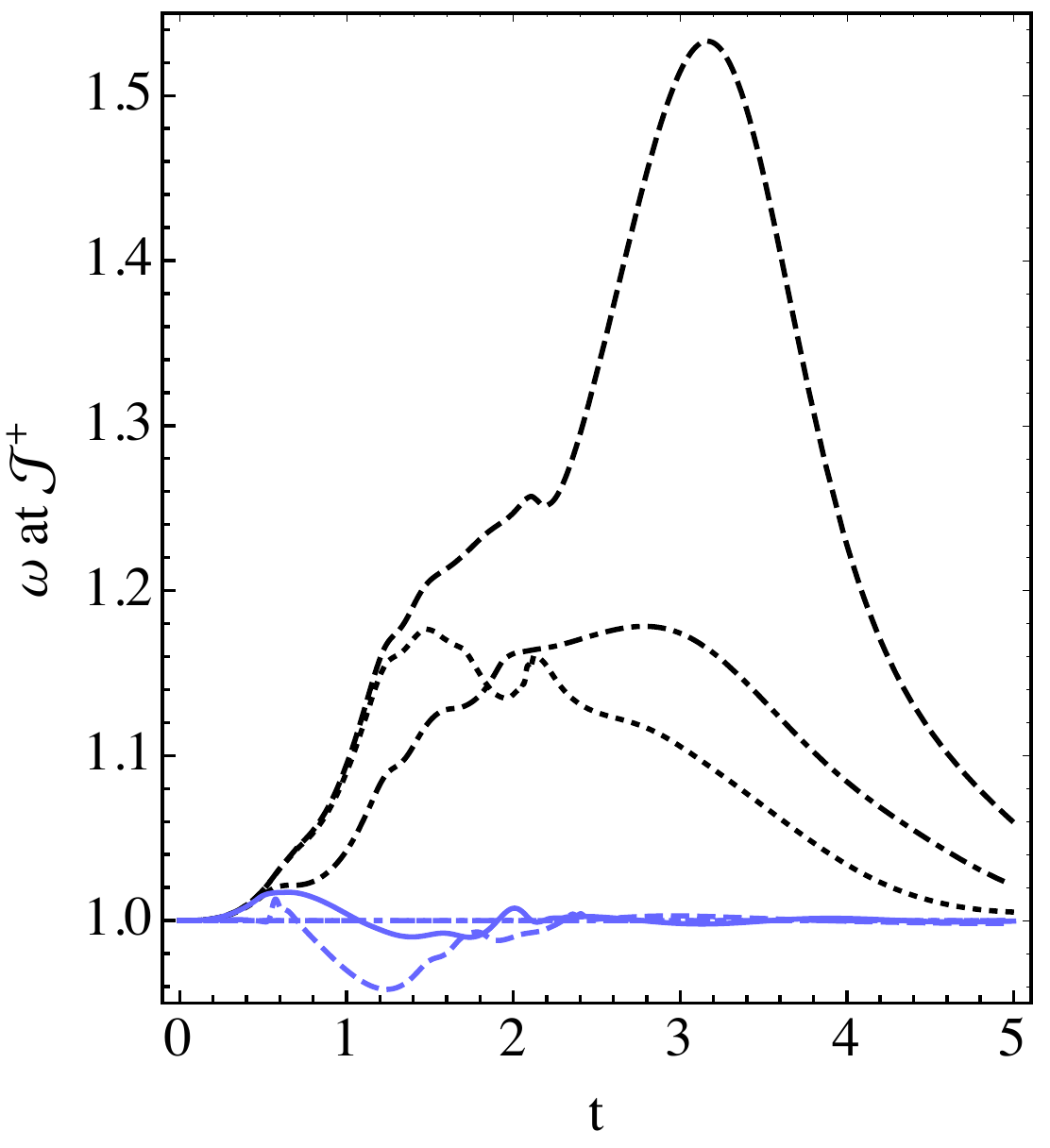}
  \includegraphics[width=0.34\linewidth]{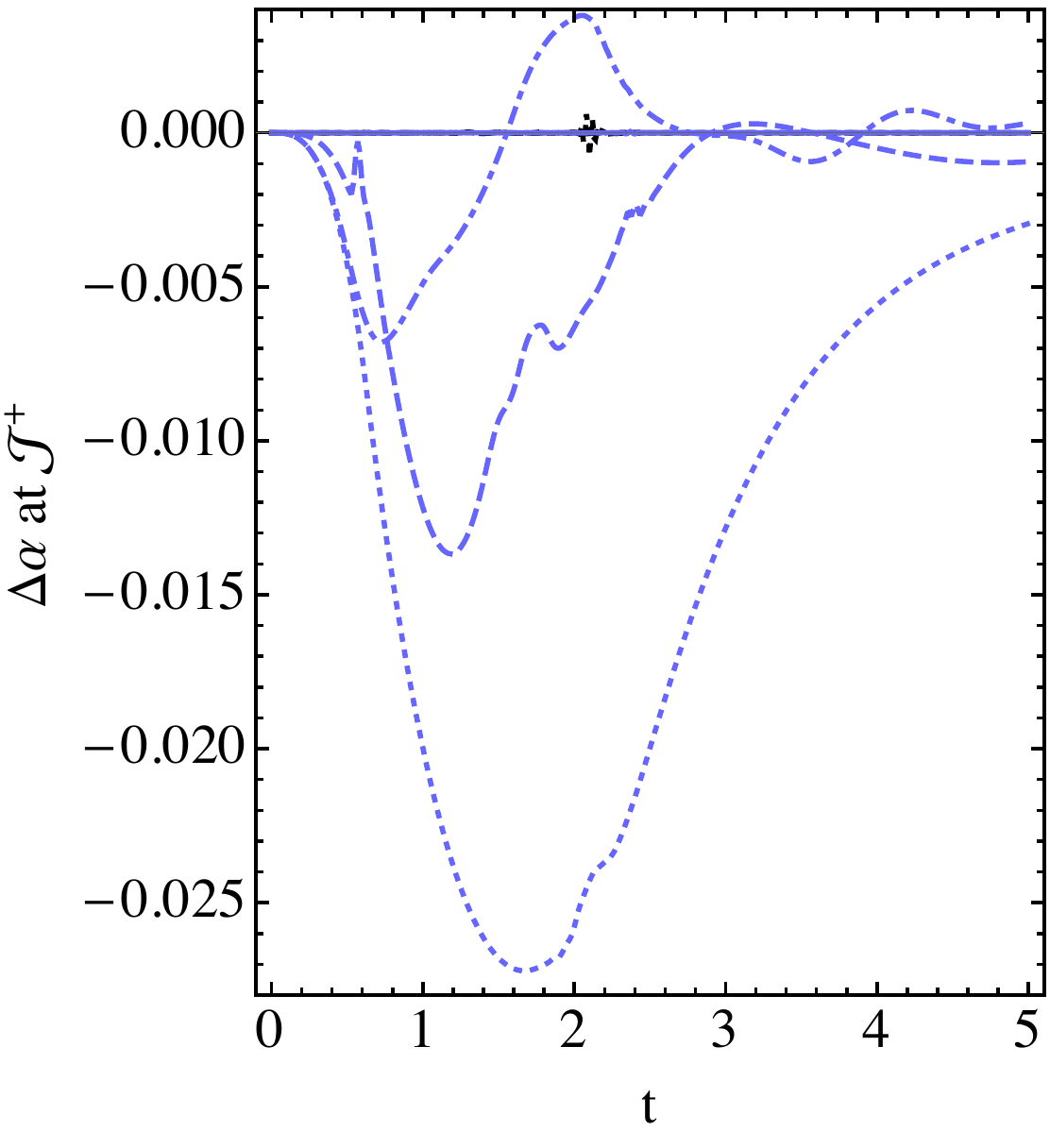}&
  \includegraphics[width=0.34\linewidth]{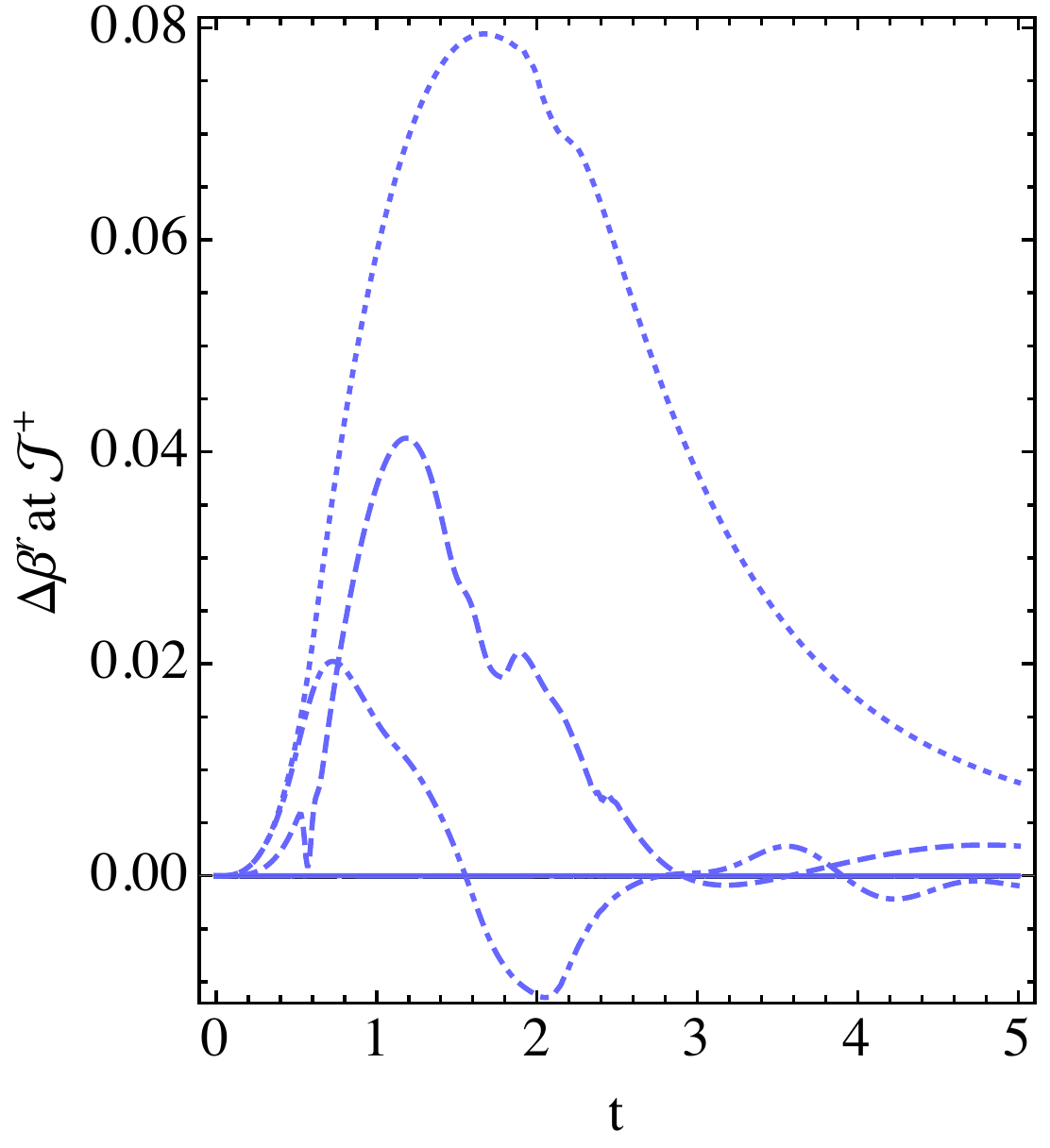}
\end{tabular}\vspace{-3ex}
\caption{Behaviour of $\atscrip{\omega}$, $\atscrip{\Delta\alpha}$ and $\atscrip{\Delta\beta^r}$ as functions of time, for the same simulations presented in \fref{f:signalsscri} and following the same legend code.
}\label{f:othersscri}
\end{figure}

The simulations whose data are presented in figures \ref{f:signalsscri} and \ref{f:othersscri} used a staggered grid, which means that the curves shown correspond to data extrapolated to $\scri^+$ for half a spatial step.
The basic parameters were given at the beginning of this section, while the specific gauge parameters for each simulation will be indicated in the following paragraph.
As mentioned above, the simulations included here are only examples of the vast range of possible configurations.


Figure \ref{f:signalsscri} shows the signal of the rescaled scalar field $\bar\Phi=\tilde\Phi/\Omega$ at future null infinity as raw output from the code. The exception is the black solid line: it corresponds to the signal rescaled by $\hat\Omega$, the conformal factor that satisfies the preferred conformal gauge, and is plotted against the Bondi time coordinate.
The following 3 signals listed in the legend were produced in simulations with fixed (time-independent) shift and i) harmonic lapse, ii) cK condition with $m_{cK}=1$, and iii) cK condition with $m_{cK}=1$ superposed as in \eref{e:supermatchlapse} (with $n_{\alpha}=2$) on the harmonic condition.
The 5th listed signal corresponds to a simulation with 1+log ($n_{1+log}=1$) and integrated Gamma-driver ($\lambda=3/4$ and $\xi_{\beta^r}=2$, so that the shift is forced to remain fixed at $\scri^+$, see right plot in \fref{f:othersscri}), without any matching.
The 6th curve was obtained from a simulation with harmonic lapse and shift, with only a small superposition \eref{e:supermatchshift} (with $n_{\beta^r}=2$ and $\lambda=0.1$) to keep the shift condition stable at the origin without adding extra damping.
The final two signals correspond to simulations where the preferred conformal gauge is satisfied at $\scri^+$.
Both include a matching of the form \eref{eg:matchf2} in \eref{e:matchinout} with the cK condition ($n_{cK}=0.4$) and the integrated Gamma-driver ($\lambda=1$): in the first one the matching is performed practically at the origin ($r_-=0$ and $r_+=0.05$, this is why it is labelled simply as preferred conformal gauge), while the second one uses $r_-=0.4$ and $r_+=0.9$. For these two simulations $\atscrip{\Omega}=\atscrip{\hat\Omega}$ holds (see how $\atscrip{\omega}=1$ at all times on the left in \fref{f:othersscri}) and thus the amplitude of the scalar field signal is the same as for the black solid line.

Other relevant quantities are plotted at $\scri^+$ in \fref{f:othersscri}: the auxiliary conformal factor $\omega=\hat\Omega/\Omega$ and the variations of the lapse and shift, $\Delta\alpha=\alpha-\backalpha$ and $\Delta\beta^r=\beta^r-\backbetar$.
The values of $\atscrip{\omega}$ differ from unity for the first five simulations, where the preferred conformal gauge does not hold.
The lapse and shift stay fixed at $\scri^+$ in the first four simulations (the lapse is ``damped'' in all of them, while the shift is fixed in the first three ones and ``damped'' in the fourth one), and so ensure that no incoming modes appear at $\scri^+$. They are however allowed to move in the simulations with harmonic-like behaviour at $\scri^+$: harmonic lapse and shift, and the two conditions (the one matched near the origin and the one matched in the middle of the domain) that satisfy the preferred conformal gauge. Snapshots at constant time of $\alpha(r)$ and $\Delta\beta^r(r)$ for similar gauge conditions are presented in figure 8.9 in \cite{\alexthesis}.

It is evident from \fref{f:signalsscri} that an evolved shift plays a very important role: its effect in controlling the behaviour of $\Lambda^r$ causes less distortion in the spacetime and the evolved quantities reach their final states earlier.
The left plot in \fref{f:othersscri} also shows that $\atscrip{\omega}$'s deviation from unity is much smaller than in the cases with time-independent shift.
Of the two cases that satisfy the preferred conformal gauge at $\scri^+$, the one that is matched to a larger interior region with cK slicing condition and Gamma-driver (last listed case in legend) shows smaller deviations in lapse and shift at $\scri^+$ and a scalar field signal at $\scri^+$ closer to the black solid line. It is possible that larger characteristic speeds in the interior or stronger couplings with the $K$ and $\Lambda^r$ variables decrease the spacetime distortion caused by the perturbation more effectively.
The signal from the simulation with unmatched 1+log and integrated Gamma-driver, not one of the best options due to its potential non-vanishing incoming modes, is surprisingly very close to the undistorted signal measured with the Bondi time coordinate - at least around the minimum in the signal, which corresponds to the part of the scalar field perturbation that was reflected at the origin (see figure 5 in \cite{\pap}).
This is a combination of the ``damped'' lapse and shift at $\scri^+$ and the fact that the auxiliary conformal factor $\omega$ at $\scri^+$ is practically one again when the reflected signal leaves the domain through $\scri^+$ (the similarity with the undistorted ``Bondi'' signal is probably not as high in the first peak).
This behaviour is not completely by chance, as the same gauge conditions with slightly different parameter choices give a similar result.
However, changing the initial perturbation may provide different results.

\section{Simulations with a non-staggered grid: point on $\scri^+$}\label{s:ptscri}

Here we will describe the implementation of the relevant technical improvement that allows us to include $\scri^+$ on a gridpoint in our numerical domain.

\subsection{Staggered vs. non-staggered grid}

The Einstein equations expressed in terms of the conformally rescaled metric $\bar g_{ab}$ and reduced to spherical symmetry formally diverge at two points of the compactified radial coordinate. The first one is the coordinate origin $r=0$, that is just a coordinate singularity for regular data but in presence of a black hole corresponds to a compactified asymptotically flat end for wormhole data (expressed as a puncture \cite{Brandt:1997tf}) or to the infinitely far end of a trumpet \cite{Hannam:2006vv} for trumpet data. The other diverging point is the location of future null infinity $r=\rscri$, where the conformal factor $\Omega$ vanishes.

\begin{figure}[htbp]
\center
\begin{tabular}{cc}
	\begin{tikzpicture}[scale=0.65]
		\draw[dashed] (10.5cm, 0cm) -- (11.5cm, 0cm); \draw (11.5cm, 0cm) -- (22cm, 0cm);
		\draw (18cm, 70pt) -- (18cm, - 10pt); \draw (18cm, 0cm) node[above=52pt] {$r=\rscri$};
		\foreach \x in {12, 14, 16} \fill (\x cm, 0cm) circle (4.5pt);
		\foreach \x in {20, 22} \draw (\x cm, 0cm) circle (4.5pt);
		\foreach \x in {14,16,18,20,22} \draw[dashed] (\x cm,0cm) -- (18cm, 2cm);
		\foreach \x in {12, 14, 16} \fill (\x cm, 2cm) circle (4.5pt);
    \foreach \x in {18} \fill[gray] (\x cm, 0cm) circle (4.5pt);
    \fill[gray] (18cm, 2cm) circle (4.5pt);
	\end{tikzpicture}
	\begin{tikzpicture}[scale=0.65]
		\draw[dashed] (10.5cm, 0cm) -- (11.5cm, 0cm); \draw (11.5cm, 0cm) -- (22cm, 0cm);
		\draw (19cm, 70pt) -- (19cm, - 10pt); \draw (19cm, 0cm) node[above=52pt] {$r=\rscri$};
		\foreach \x in {12, 14, 16, 18} \fill (\x cm, 0cm) circle (4.5pt);
		\foreach \x in {20, 22} \draw (\x cm, 0cm) circle (4.5pt);
		\fill (18cm, 2cm) circle (4.5pt); \foreach \x in {12,14,16,18,20} \draw[dashed] (\x cm,0cm) -- (18cm, 2cm);
		\foreach \x in {12, 14, 16} \fill (\x cm, 2cm) circle (4.5pt);
	\end{tikzpicture}
\end{tabular}
	\caption{Non-staggered grid on the left and staggered one on the right, at the location of $\scri^+$ in the domain. The values of the variables on the black-filled points are evolved using the equations of motion and the empty circles denote ghost points, which are required to calculate the derivatives at the boundaries and are filled using the boundary conditions. The gray-filled points represent the location of the grid where the equations are formally singular and a special treatment is required. The dashed lines on the left indicate a 4th order centered stencil, while those on the right represent a 4th order one-point off-centered upwind stencil.} \label{grids}
\end{figure}
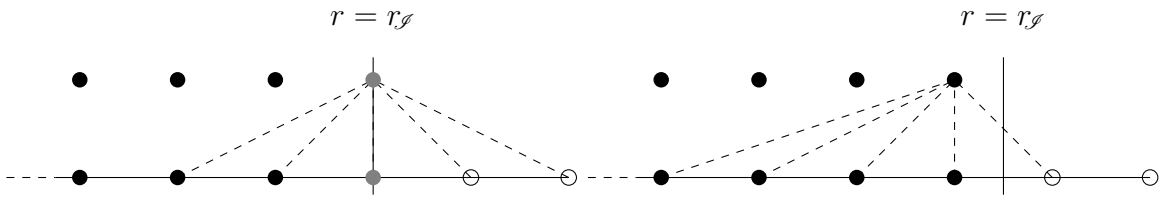

The results presented in \cite{\pap,\procere,\procmg,\alexthesis} were obtained using a staggered grid in our numerical implementation. In a staggered grid, the gridpoints where the variables and equations are evaluated during evolution do not coincide with any of the locations where the equations diverge ($r=0$ and $r=\rscri$), as illustrated on the right diagram in \fref{grids}. This simplifies the implementation considerably, because no special treatment of the RHSs on the divergent points is required. However, the value of the quantities at $\scri^+$ is only obtained by extrapolation (for less than a spatial step, half a step in our case).

The non-staggered grid in our setup includes the radial locations $r=0$ and $r=\rscri$ as actual gridpoints (see diagram on the left of \fref{grids}) and allows us to directly calculate our quantities at $\scri^+$, but its implementation is considerably more difficult: the formally diverging RHSs of the spherically symmetric evolution equations (in Appendix C of \cite{\pap} and in Chapter 2 of \cite{\alexthesis}) have to be transformed so that a finite numerical value is obtained at the diverging points. The treatment of $r=0$ is an important issue: a non-staggered grid at the origin can be suitable for the evolution of regular spacetimes (using the parity behaviour of the evolution variables to determine their values at $r=0$\footnote{Odd variables will necessarily vanish there, while even ones have a vanishing first derivative at $r=0$, which can be used to calculate their value there. More details are included in subsection 6.3.2 in \cite{\alexthesis}.}), but a staggered origin is a better choice for evolving a black hole. For instance, in the moving puncture approach \cite{Bruegmann:1997uc,Bruegmann:2006at} the location of the puncture that represents each of the black holes must never coincide with any of the gridpoints. Therefore, when including a black hole in the evolved spacetime \cite{\papbh}, we will use a mixed grid, where the origin $r=0$ is staggered and $\scri^+$ is located on an actual gridpoint.

\subsection{Implementation of the equations for the non-staggered grid at $\scri^+$}\label{ss:nostagimplem}

The modification of our spherically symmetric reduction of the equations of motion for their evaluation at $\scri^+$ is performed in two steps.
First, note that the terms divided by $\Omega$ (and $\Omega^2$) that appear in the RHSs are expected to cancel and provide finite RHSs that attain a finite value at $r=\rscri$ - Friedrich showed that the Einstein equations expressed as the Conformal Field Equations attain a regular limit at $\scri^+$ \cite{friedrich1983,Friedrich:2003fq} for the type of regular initial data we are interested in. This means that for these kind of solutions the sum of the numerators of the diverging terms in our equations (in Appendix C of \cite{\pap} and (2.82) in Chapter 2 of \cite{\alexthesis}) must necessarily vanish at $\scri^+$ at the appropriate order.
These relations (numerators = 0) is what we call the regularity conditions at $\scri^+$, which for our formulation and reduction of the Einstein equations are given by \eref{e:regconds}, and they will be used to calculate the regular limit of the RHSs at $\scri^+$.

\subsubsection{Impose the regularity conditions on the variables at $\scri^+$:}\label{ss:regconds}

The regularity conditions for the conformally rescaled Einstein equations that we use in our code (included in Chapter 2 of \cite{\alexthesis} - note that the metric coefficient $\gamma_{\theta\theta}\equiv\gamma_{rr}^{-1/2}$ is eliminated in the equations implemented in our code) assuming that the preferred conformal gauge holds at future null infinity are the following:
\begin{subequations}\label{e:regconds}
\begin{eqnarray}
\atscrip{\chi}&=& \atscrip{\gamma_{\theta\theta}} \equiv \atscrip{\gamma_{rr}^{-1/2}} , \label{e:regcond1} \\
\atscrip{\Delta \tilde K}&=& \atscrip{\left(-\Kc - \frac{3\beta^r\Omega'}{\alpha}\right)} , \\
\atscrip{\alpha}&=& \atscrip{-\beta^r\sqrt{\frac{\gamma_{rr}}{\chi}}} \quad (\textrm{comes from } \atscrip{\bar g_{tt}}=0) , \label{e:regcond3} \\
\atscrip{\tilde \Theta}&=&0 ,
\end{eqnarray}
\end{subequations}
as well as two much lengthier expressions for $\atscrip{A_{rr}}$ and $\atscrip{\Lambda^r}$, obtained by solving for these quantities the numerators of the $\Omega^{-1}$ terms in $\dot A_{rr}$ and $\dot{\Lambda^r}$. The equation of motion of the massless scalar field does not introduce any extra regularity conditions.

The regularity condition that arises from the divergent terms in the gauge conditions that satisfy the preferred conformal gauge (described in \sref{s:pref}) is
\begin{eqnarray}\label{e:regbetar}
\atscrip{\beta^r}&=& \atscrip{\left[\frac{3\alpha^2}{\Kc \ \rscri} +\cpbg\left( \frac{\Kc \ \rscri \  \alpha^2}{6\,\Omega'} + \frac{\alpha^3}{2\,\Omega'}\right) \right]}.
\end{eqnarray}
Substituting the value of $\atscrip{\beta^r}$ according to \eref{e:regbetar} into \eref{e:afinet}, using also $\atscrip{\Omega'}=\Kc\case{\rscri}{3}$ and $\cpbg=0$ yields that the code time $t$ at future infinity equals the Bondi time $t_B$. Unfortunately, at the current state our simulations are not stable with a vanishing $\cpbg$.

We implemented the regularity conditions in the code in the following way: as there is only one degree of freedom available at $\scri^+$ for our spherically symmetric reduction of the Einstein equations (8 evolution variables - $\chi$, $\gamma_{rr}$, $\alpha$, $\beta^r$, $A_{rr}$, $\Delta \tilde K$, $\Lambda^r$ and $\tilde \Theta$ - and only 7 regularity conditions), we choose to identify it with the free value of $\gamma_{rr}$. Using relations \eref{e:regconds} and \eref{e:regbetar}, we express all regularity conditions in terms of $\gamma_{rr}$ (this may not be a requirement for stability) and the derivatives of the other variables.
Note that the regularity conditions for lapse and shift as expressed above are circular: in order to obtain the expressions to impose in the code we substitute $\chi$ in \eref{e:regcond3} in terms of $\gamma_{rr}$ as in \eref{e:regcond1}, and then solve \eref{e:regcond3} and \eref{e:regbetar} for $\atscrip{\alpha}$ and $\atscrip{\beta^r}$. The correct result for $\cpbg\neq0$ is given by
\begin{subequations}
\begin{eqnarray}
\fl \atscrip{\beta^r} &= \atscrip{-\frac{\alpha}{\gamma_{rr}^{3/2}}} , \\
\fl \atscrip{\alpha} &= \left(-\frac{\Kc \ \rscri}{6}-\frac{3 \Omega'}{\cpbg \Kc \ \rscri}+\atscrip{\frac{\sqrt{\left(-\frac{\cpbg}{3}  \Kc \ \rscri-\frac{6 \Omega '}{\Kc \ \rscri}\right)^2-\frac{8 \cpbg \Omega '}{\gamma_{rr}^{3/4}}}}{2 \cpbg}\right)}, 
\end{eqnarray}
\end{subequations}
while for $\cpbg=0$ we obtain the simple relations
\begin{equation}
  \atscrip{\beta^r} = \atscrip{\frac{\Kc\ \rscri}{3\gamma_{rr}^{3/2}}} \quad \textrm{and} \quad \atscrip{\alpha} = \atscrip{-\frac{\Kc\ \rscri}{3\gamma_{rr}^{3/4}}} .
\end{equation}
\noindent The final expressions are used to impose the values of the evolution variables at the gridpoint corresponding to $\scri^+$ (except for $\gamma_{rr}$, which is determined by its equation of motion) after each time step.

\subsubsection{Apply l'H\^opital rule to the divergent RHSs at $\scri^+$:}\label{ss:lhopit}

Due to their considerable length, we are not including the equations of motion that are implemented numerically at $\scri^+$ in our code, but we describe how to calculate them.
The following equation is taken as an example: variable X's RHS has a regular part at $\scri^+$, denoted by A (it depends on other evolution variables and on $r$), as well as two formally singular terms at $\scri^+$ (those divided by a power of $\Omega$):
\begin{equation}
\dot X = A + \frac{B}{\Omega} -\frac{C}{\Omega} \quad (\textrm{with regularity condition }\atscrip{B}=\atscrip{C}) .
\end{equation}
The l'H\^opital rule is applied on the diverging terms as
\begin{equation}
\lim_{r\to\rscri}\frac{B-C}{\Omega}  = \frac{0}{0} \equiv \lim_{r\to\rscri}\frac{(B-C)'}{\Omega'} = \atscrip{\left[\frac{(B-C)'}{\Omega'}\right]} ,
\end{equation}
where the last expression is regular at future null infinity. This limit is substituted in the RHS to be evaluated at $\scri^+$ as
\begin{equation}
\atscrip{\dot X} = \atscrip{A} +\atscrip{\left(\frac{B}{\Omega} -\frac{C}{\Omega}\right)} \to \quad\atscrip{\dot X} = \atscrip{A}  + \atscrip{\left[\frac{(B-C)'}{\Omega'}\right]} .
\end{equation}
This procedure is followed for each of the diverging RHSs.
In the presence of terms that diverge as $\Omega^{-2}$ the l'H\^opital rule is applied twice:
\begin{eqnarray}
\fl \left\{\atscrip{\dot Y} =\right.& \left.\atscrip{D} + \atscrip{\left(\frac{E}{\Omega} +\frac{F}{\Omega^2}\right)}\right\} \cdot\Omega \ \to\  0=\atscrip{E}+\atscrip{\left(\frac{F}{\Omega}\right)}\ \to \ \atscrip{F}=0 , \nonumber \\
\fl \atscrip{\dot Y} =& \atscrip{D} +\atscrip{\left(\frac{E}{\Omega} +\frac{F'/\Omega'}{\Omega}\right)} \ \to \ \atscrip{\left(E+F'/\Omega'\right)}=0, \nonumber \\
\fl \atscrip{\dot Y} =& \atscrip{D} +\atscrip{\left[\frac{(E+F'/\Omega')'}{\Omega'}\right]}
\end{eqnarray}
The regularity conditions \eref{e:regconds} (as well as those for $\atscrip{A_{rr}}$ and $\atscrip{\Lambda^r}$) and \eref{e:regbetar} are substituted into the ``l'H\^opitalized'' RHSs at the analytical level.

\subsection{Results and comparison to staggered grid}

\begin{figure}[htbp!!]
\center
\includegraphics[width=1.00\linewidth]{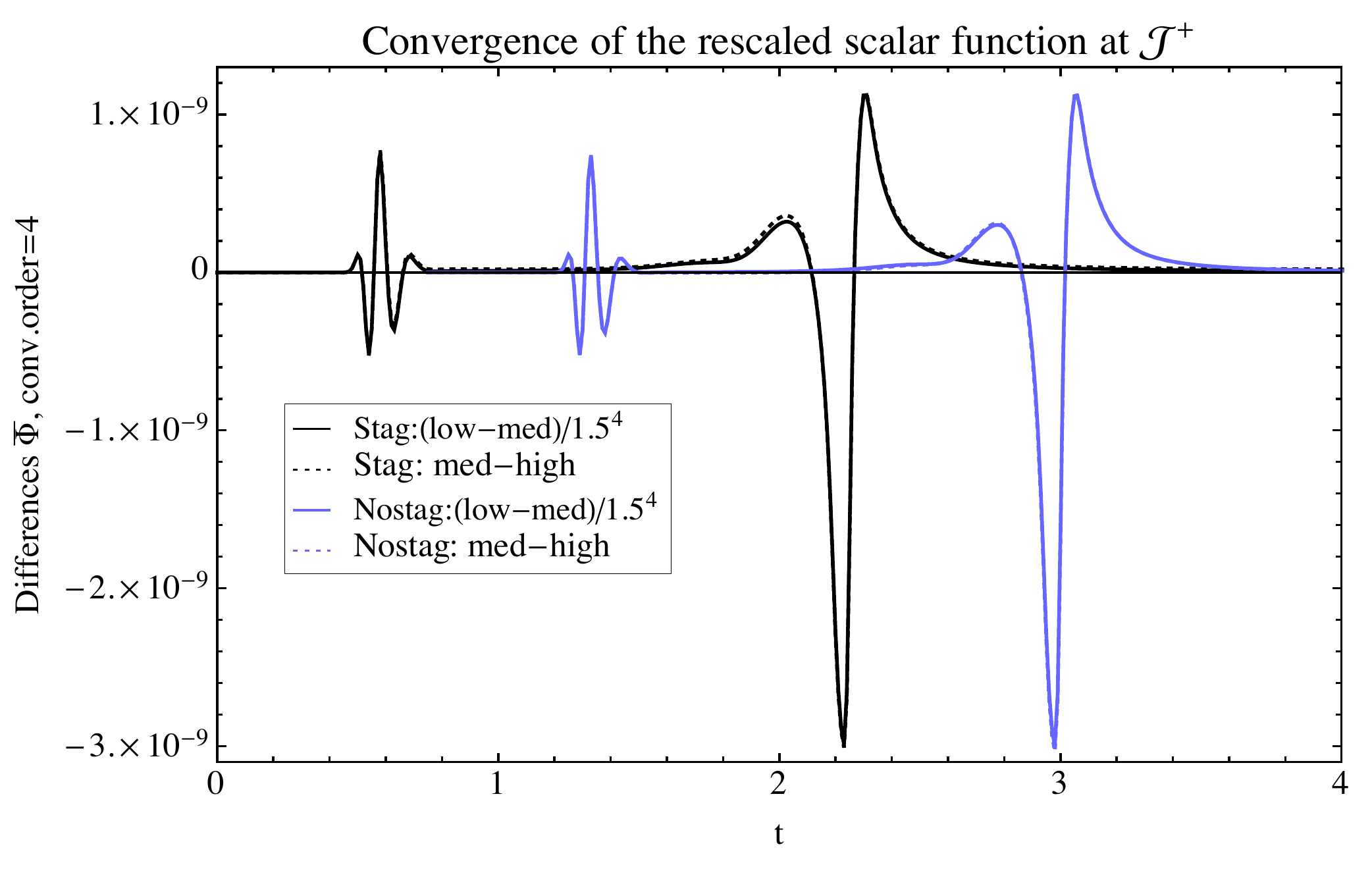}
\caption{4th order convergence of the rescaled scalar field $\tilde\Phi/\Omega$ at $\scri^+$, comparing the results with a staggered (extrapolated half a spatial step) and a non-staggered (evaluated on $\scri^+$) grids.}
\label{fr:convpointscri}
\end{figure}

Comparison of the performance of the staggered and non-staggered grids at $\scri^+$ is shown in \fref{fr:convpointscri} as a convergence test.
The corresponding simulations used the same initial data as described in \sref{s:signals}.
The Z4c formulation as well as the gauge conditions satisfying the preferred conformal gauge (\sref{s:pbg} and \ref{a:pbg}) were used.
The curves correspond to the differences between the signals at $\scri^+$ in simulations with 1600 (low resolution), 2400 (medium) and 3600 (high) points. The non-staggered data (in blue/gray) has been shifted 0.75 forward in time to be able to compare both sets of data in a clearer way.
The expected 4th order convergence is very good in both cases, even if it looks slightly better in the non-staggered one.

\section{Conclusions}

In \cite{\pap} we presented our first approach to the hyperboloidal initial value problem in spherical symmetry, describing a very basic setup in terms of gauge conditions: harmonic slicing and fixed shift.
The preferred conformal gauge was not satisfied, which made an affine reparameterization of the time coordinate at $\scri^+$ more cumbersome, and future null infinity was not included in the numerical grid.

As part of the development of a robust framework to solve the hyperboloidal initial value problem, here we have successfully implemented the preferred conformal gauge in the form of hyperbolic gauge conditions \eref{eg:harmtwobp} that enforce \eref{e:prefconfgauge}, without altering our time-independent conformal factor $\Omega$ \eref{ein:omega} and the simplifications that the latter adds to our setup.
These gauge conditions were also designed to systematically include appropriate source functions, required for the hyperboloidal evolution due to its non-trivial background.
In our current implementation the time coordinate at $\scri^+$ is related \eref{e:afinet} in a non-trivial way to the Bondi time. The condition \eref{extrarel} that makes them equal is satisfied with the parameter choice $\cpbg=0$, but this is not yet achievable in our simulations for stability reasons. Thus, for the moment our code does not directly provide an affine time coordinate at $\scri^+$ at all times.
However, the preferred conformal gauge makes the ``a posteriori'' calculation of the Bondi time much easier, as compared to the simpler setup in \cite{\pap}.

The study of the different slicing and shift conditions presented in this paper has increased our understanding of how to adapt current common options of hyperbolic gauge conditions (like the Bona-Mass\'o family or the Gamma-driver) to the free evolution of the hyperboloidal initial value problem.
The main ingredients are still those presented in \cite{\pap}: the addition of damping terms and source functions.
Rather than in the region near $\scri^+$, the tested slicing and shift conditions show the most representative differences in behaviour in the interior part of the domain, towards the origin. This effect will be more relevant in the treatment of spacetimes that include a black hole or very large perturbations of regular spacetimes, and will be described in \cite{\papbh}.
Our current setup, where we choose not to have incoming gauge modes at future null infinity, requires physical characteristic speeds at $\scri^+$, so that the matching (using \eref{e:superpmatch} or \eref{e:matchinout}) of at least some coefficients in the gauge conditions is necessary. Where in the integration domain and how this matching is performed is still an open question.
At the current stage it is difficult to determine which gauge conditions are preferable, as a change in some of the input parameters, the type of initial data (regular or including a black hole) or the used formulation of the Einstein equations can have an important impact on the performance of different slicing and shift conditions.
To exemplify the influence that gauge conditions have on the signals extracted at future null infinity, we compared the values of the rescaled scalar field at $\scri^+$ for several combinations of slicing and shift conditions and included the undistorted signal expressed in terms of the Bondi time in \fref{f:signalsscri}. These examples served to illustrate and understand the differences in behaviour of the presented gauge choices: the main conclusions are that both an evolved shift, as opposed to a fixed time-independent shift, and larger characteristic speeds in the inner part of the domain (towards the origin) cause considerably less distortion at $\scri^+$.
We hope that the different lapse and shift equations compiled here will start to conform a testbed of hyperbolic gauge conditions to study the hyperboloidal problem, and that the insights gained will be useful to future attempts at solving the problem in 3 spatial dimensions.

We have also added a relevant technical improvement to our implementation: the inclusion of a point on $\scri^+$ in the integration domain.
This has greatly benefited from the cancellations introduced by the preferred conformal gauge \eref{e:prefconfgauge}, although a gridpoint on $\scri^+$ should also be feasible without this choice.
The point on $\scri^+$ represents a step forward in both our understanding of the behaviour of our formulation of the Einstein equations at future null infinity and the degrees of freedom available there: only one of the evolution variables can be freely specified at $\scri^+$. Evaluating the equations on $\scri^+$ requires the adaptation of the divergent terms in the RHSs of the equations of motion as described in subsection \ref{ss:lhopit}, but extrapolation is no longer needed to evaluate the evolved quantities at future null infinity.
We have obtained consistent confirmation (see \fref{fr:convpointscri}) that our previous results with a staggered grid were already very good, although the non-staggered grid performs even better.
Having a gridpoint on $\scri^+$ will allow us to experiment with different boundary conditions for super-luminal gauge speeds and even to implement a reflecting-like boundary to perform simulations in asymptotically Anti-deSitter spacetimes.

\ack

We thank David Hilditch for valuable comments on the manuscript.
AV was supported by the FPU-grant AP2010-1697 and partially by the European Research Council Consolidator Grant 647839.
AV and SH acknowledge the support of the European Union FEDER founds, the MINECO grants FPA2010-16495, FPA2013-41042-P and CSD2009-00064 of the Spanish Ministry of Economy and Competitiveness and the Conselleria d'Economia i Competitivitat of the Govern de les Illes Balears.
Most of the algebraic derivations were performed using the Mathematica package xAct \cite{xAct} and some of the numerical simulations took advantage of computational resources provided by the STFC grant ST/I006285/1. 

\appendix

\section{Conformally compactified equations, notation and variables}\label{s:eqs}

In our simulations we use either the Generalized BSSN formulation \cite{PhysRevD.52.5428,Baumgarte:1998te,Brown:2007nt} or a similar conformal version of the Z4 formulation \cite{bona-2003-67,Alic:2011gg,Sanchis-Gual:2014nha}, the \CZ{} equations \cite{Bernuzzi:2009ex,Weyhausen:2011cg}, in their spherically symmetric reduction. The derivation of the equations for the conformally rescaled metric is described in \cite{\pap} and in Chapter 2 of \cite{\alexthesis} and the actual equations used in the simulations can be found in Appendix C of \cite{\pap} and again in Chapter 2 of \cite{\alexthesis}.
Our notation for the metrics is the same one as used in \cite{\pap}: the 4-dimensional physical metric is denoted as $\tilde g$, the 4d-conformal metric as $\bar g$, the 3-dimensional conformal metric (induced by $\bar g$) as $\bar \gamma$, the 3d twice conformal metric as $\gamma$ and the 3d twice conformal background metric as $\hat \gamma$.
The massless scalar field $\tilde\Phi$ satisfies the wave equation in the physical spacetime, and with respect to the conformally rescaled metric it satisfies
\begin{equation}
\bar g^{\mu\nu}\bar \nabla_\mu\bar \nabla_\nu\tilde\Phi-2\bar g^{\mu\nu}\bar \nabla_\mu\tilde\Phi\frac{\bar \nabla_\nu\Omega}{\Omega}=0 .
\end{equation}

The dynamical variables are the conformally rescaled spatial metric $\gamma_{ab}=\chi\bar \gamma_{ab}$, where $\bar \gamma_{ab}$ is the spatial metric induced from $\bar g_{ab}$, and the spatial conformal factor $\chi$.
From the decomposition of the conformal extrinsic curvature tensor $\bar K_{ab}$ we evolve its conformal trace-free part $A_{ab}=\chi\bar K_{ab}-\frac{1}{3}\gamma_{ab}\bar K$ (with $\bar K=\bar K_{ab}\bar\gamma^{ab}\equiv K_{ab}\gamma^{ab}$) and its physical trace, mixed with the physical Z4 variable $\tilde \Theta$, $\tilde K = \Omega\bar K-\case{3\beta^a\partial_a\Omega}{\alpha}-2\tilde\Theta$. The actual evolution variable in the code is a variation with respect to its initial value $\Delta\tilde{K}=\tilde K-\tilde K_0=\tilde K-\Kc$ (this last parameter corresponds to the value chosen for $\tilde K$ on the initial hyperboloidal constant-mean-curvature (CMC) slice and in our notation it will be a negative value). The Z4 quantity $\tilde \Theta$ is evolved as well.
The evolution of the Z4 variable $Z_a$ is included into that of the vector $\Lambda^a=\gamma^{bc}\left(\Gamma^a_{bc}-\hat\Gamma^a_{bc}\right) +2\gamma^{ab}Z_b$, where $\Gamma^a_{bc}$ are the Christoffel symbols calculated from $\gamma_{ab}$ and $\hat\Gamma^a_{bc}$ the ones built from a time-independent background metric $\hat \gamma_{ab}$.
The evolved gauge variables are the conformal lapse $\alpha$, the shift $\beta^r$ and, if required, its auxiliary variable $B^a$.
The scalar field is expressed as $\bar\Phi=\tilde\Phi/\Omega$, together with its auxiliary variable $\bar\Pi=\tilde\Pi/\Omega=(\partial_t\tilde\Phi)/\Omega$, mainly because these quantities do not vanish at $\scri^+$ and so are more convenient for visualization purposes.

Our spherically symmetric ansatz for the line element is
\begin{equation}\label{e:linel}
\fl d\bar s^2 = - \left(\alpha^2-\chi^{-1}\gamma_{rr}{\beta^r}^2\right) dt^2 + \chi^{-1}\left[2\, \gamma_{rr}\beta^r dt\,dr +  \gamma_{rr}\, dr^2 +  \gamma_{\theta\theta}\, r^2\, d\sigma^2\right] .
\end{equation}
The freedom introduced by the spatial conformal factor $\chi$ is fixed by eliminating $\gamma_{\theta\theta}=\gamma_{rr}^{-1/2}$ in the equations implemented in the code.
After explicitly imposing the trace-freeness condition of $A_{ab}$ and substituting \eref{e:linel}, the only remaining independent component of $A_{ab}$ is $A_{rr}$. The only non-vanishing component of the quantities $\Lambda^a$, $\beta^a$, $B^a$ and $Z_a$ is the radial one, denoted respectively by $\Lambda^r$, $\beta^r$, $B^r$ and $Z_r$.
Primes denote derivatives with respect to the compactified radial coordinate $r$ and dots denote time derivatives.

\section{Gauge conditions for the preferred conformal gauge}\label{a:pbg}

The decomposition of \eref{eg:gharmpb} into lapse and shift evolution equations is
{\small
\begingroup
\begin{subequations}\label{eg:harmtwobp}
\begin{eqnarray}
\fl \dot\alpha =& {\beta^r} \alpha '-\alpha  {\beta^r}'-\frac{\alpha  {\beta^r} \gamma _{\theta \theta }'}{\gamma _{\theta \theta }}+\frac{3 \alpha  {\beta^r} \chi '}{2 \chi }+\frac{4 \alpha  {\beta^r} \Omega '}{\Omega }-\frac{2 \alpha  {\beta^r} \hat{\alpha }'}{\hat{\alpha }}+\frac{\alpha  \dot{\gamma _{\theta \theta }}}{\gamma _{\theta \theta }}-\frac{3 \alpha  \dot{\chi }}{2 \chi }-\frac{\alpha ^3 \tilde F^t}{\Omega ^2}
\nonumber \\* \fl &
-\frac{2 \alpha ^3 \chi  \hat{\gamma _{\theta \theta }} \hat{\beta ^r} \Omega '}{\hat{\alpha }^2 \hat{\chi } \Omega  \gamma _{\theta \theta }}-\frac{\alpha ^3 \chi  \hat{\gamma _{\theta \theta }} \hat{\beta ^r} \hat{\chi }'}{\hat{\alpha }^2 \hat{\chi }^2 \gamma _{\theta \theta }}+\frac{\alpha ^3 \chi  \hat{\beta ^r} \hat{\gamma _{\theta \theta }}'}{\hat{\alpha }^2 \hat{\chi } \gamma _{\theta \theta }}+\frac{2 \alpha ^3 \chi  \hat{\gamma _{\theta \theta }} \hat{\beta ^r}}{\hat{\alpha }^2 r \hat{\chi } \gamma _{\theta \theta }}-\frac{\alpha  \hat{\beta ^r} \Omega '}{\Omega }+\frac{\alpha  \hat{\alpha }' \hat{\beta ^r}}{\hat{\alpha }}
\nonumber \\* \fl &
-\frac{2 \alpha  {\beta^r}}{r}-\frac{\alpha ^3 \chi  \hat{\beta ^r} \hat{\gamma _{rr}} \Omega '}{\hat{\alpha }^2 \hat{\chi } \Omega  \gamma _{rr}}+\frac{\alpha ^3 \chi  \hat{\gamma _{rr}} \hat{\beta ^r}'}{\hat{\alpha }^2 \hat{\chi } \gamma _{rr}}-\frac{\alpha ^3 \chi  \hat{\beta ^r} \hat{\gamma _{rr}} \hat{\chi }'}{2 \hat{\alpha }^2 \hat{\chi }^2 \gamma _{rr}}+\frac{\alpha ^3 \chi  \hat{\beta ^r} \hat{\gamma _{rr}}'}{2 \hat{\alpha }^2 \hat{\chi } \gamma _{rr}}+\frac{\alpha  {\beta^r}^2 \hat{\beta ^r} \hat{\gamma _{rr}} \Omega '}{\hat{\alpha }^2 \hat{\chi } \Omega }
\nonumber \\* \fl &
-\frac{\alpha  {\beta^r}^2 \hat{\gamma _{rr}} \hat{\beta ^r}'}{\hat{\alpha }^2 \hat{\chi }}+\frac{\alpha  {\beta^r}^2 \hat{\beta ^r} \hat{\gamma _{rr}} \hat{\chi }'}{2 \hat{\alpha }^2 \hat{\chi }^2}-\frac{\alpha  {\beta^r}^2 \hat{\beta ^r} \hat{\gamma _{rr}}'}{2 \hat{\alpha }^2 \hat{\chi }}-\frac{2 \alpha  {\beta^r} \hat{\beta ^r}^2 \hat{\gamma _{rr}} \Omega '}{\hat{\alpha }^2 \hat{\chi } \Omega }+\frac{2 \alpha  {\beta^r} \hat{\beta ^r} \hat{\gamma _{rr}} \hat{\beta ^r}'}{\hat{\alpha }^2 \hat{\chi }}
\nonumber \\* \fl &
-\frac{\alpha  {\beta^r} \hat{\beta ^r}^2 \hat{\gamma _{rr}} \hat{\chi }'}{\hat{\alpha }^2 \hat{\chi }^2}+\frac{\alpha  {\beta^r} \hat{\beta ^r}^2 \hat{\gamma _{rr}}'}{\hat{\alpha }^2 \hat{\chi }}+\frac{\alpha  \hat{\beta ^r}^3 \hat{\gamma _{rr}} \Omega '}{\hat{\alpha }^2 \hat{\chi } \Omega }-\frac{\alpha  \hat{\beta ^r}^2 \hat{\gamma _{rr}} \hat{\beta ^r}'}{\hat{\alpha }^2 \hat{\chi }}+\frac{\alpha  \hat{\beta ^r}^3 \hat{\gamma _{rr}} \hat{\chi }'}{2 \hat{\alpha }^2 \hat{\chi }^2}-\frac{\alpha  \hat{\beta ^r}^3 \hat{\gamma _{rr}}'}{2 \hat{\alpha }^2 \hat{\chi }}
\nonumber \\ \fl &
-\frac{\alpha  {\beta^r} \gamma _{rr}'}{2 \gamma _{rr}}+\frac{\alpha  \dot{\gamma _{rr}}}{2 \gamma _{rr}}
 , \label{eg:harmderlapsebp} \\
\fl \dot \beta^r =& -\frac{\alpha ^2 \tilde F^r}{\Omega ^2}+\frac{2 {\beta^r}^2}{r}+\frac{{\beta^r} \dot{\alpha }}{\alpha }+\frac{3 {\beta^r} \dot{\chi }}{2 \chi }-\frac{{\beta^r}^2 \alpha '}{\alpha }-\frac{\alpha  \chi  \alpha '}{\gamma _{rr}}+2 {\beta^r} {\beta^r}'-\frac{3 {\beta^r}^2 \chi '}{2 \chi }+\frac{\alpha ^2 \chi '}{2 \gamma _{rr}}
\nonumber \\* \fl &
+\frac{2 \hat{\beta ^r}^2 \Omega '}{\Omega }-\frac{2 {\beta^r} \hat{\beta ^r} \Omega '}{\Omega }-\frac{\hat{\beta ^r}^4 \hat{\gamma _{rr}} \Omega '}{\Omega  \hat{\alpha }^2 \hat{\chi }}+\frac{2 {\beta^r} \hat{\beta ^r}^3 \hat{\gamma _{rr}} \Omega '}{\Omega  \hat{\alpha }^2 \hat{\chi }}-\frac{{\beta^r}^2 \hat{\beta ^r}^2 \hat{\gamma _{rr}} \Omega '}{\Omega  \hat{\alpha }^2 \hat{\chi }}-\frac{3 {\beta^r}^2 \Omega '}{\Omega }-\frac{\hat{\alpha }^2 \hat{\chi } \Omega '}{\Omega  \hat{\gamma _{rr}}}
\nonumber \\* \fl &
+\frac{\alpha ^2 \chi  \hat{\beta ^r}^2 \hat{\gamma _{rr}} \Omega '}{\Omega  \hat{\alpha }^2 \hat{\chi } \gamma _{rr}}+\frac{3 \alpha ^2 \chi  \Omega '}{\Omega  \gamma _{rr}}+\frac{2 \alpha ^2 \chi  \hat{\beta ^r}^2 \hat{\gamma _{\theta \theta }} \Omega '}{\Omega  \hat{\alpha }^2 \hat{\chi } \gamma _{\theta \theta }}-\frac{2 \alpha ^2 \chi  \hat{\gamma _{\theta \theta }} \Omega '}{\Omega  \hat{\gamma _{rr}} \gamma _{\theta \theta }}-\frac{\hat{\beta ^r}^2 \hat{\alpha }'}{\hat{\alpha }}+\frac{2 {\beta^r} \hat{\beta ^r} \hat{\alpha }'}{\hat{\alpha }}+\frac{\hat{\alpha } \hat{\chi } \hat{\alpha }'}{\hat{\gamma _{rr}}}
\nonumber \\* \fl &
-\hat{\beta ^r} \hat{\beta ^r}'+\frac{\hat{\beta ^r}^3 \hat{\gamma _{rr}} \hat{\beta ^r}'}{\hat{\alpha }^2 \hat{\chi }}-\frac{2 {\beta^r} \hat{\beta ^r}^2 \hat{\gamma _{rr}} \hat{\beta ^r}'}{\hat{\alpha }^2 \hat{\chi }}+\frac{{\beta^r}^2 \hat{\beta ^r} \hat{\gamma _{rr}} \hat{\beta ^r}'}{\hat{\alpha }^2 \hat{\chi }}-\frac{\alpha ^2 \chi  \hat{\beta ^r} \hat{\gamma _{rr}} \hat{\beta ^r}'}{\hat{\alpha }^2 \hat{\chi } \gamma _{rr}}-\frac{\hat{\beta ^r}^4 \hat{\gamma _{rr}} \hat{\chi }'}{2 \hat{\alpha }^2 \hat{\chi }^2}
\nonumber \\* \fl &
+\frac{{\beta^r} \hat{\beta ^r}^3 \hat{\gamma _{rr}} \hat{\chi }'}{\hat{\alpha }^2 \hat{\chi }^2}-\frac{{\beta^r}^2 \hat{\beta ^r}^2 \hat{\gamma _{rr}} \hat{\chi }'}{2 \hat{\alpha }^2 \hat{\chi }^2}-\frac{{\beta^r}^2 \hat{\chi }'}{2 \hat{\chi }}+\frac{\hat{\beta ^r}^2 \hat{\chi }'}{2 \hat{\chi }}+\frac{\alpha ^2 \chi  \hat{\beta ^r}^2 \hat{\gamma _{rr}} \hat{\chi }'}{2 \hat{\alpha }^2 \hat{\chi }^2 \gamma _{rr}}+\frac{\alpha ^2 \chi  \hat{\chi }'}{2 \hat{\chi } \gamma _{rr}}-\frac{\alpha ^2 \chi  \hat{\gamma _{\theta \theta }} \hat{\chi }'}{\hat{\chi } \hat{\gamma _{rr}} \gamma _{\theta \theta }}
\nonumber \\* \fl &
+\frac{\alpha ^2 \chi  \hat{\beta ^r}^2 \hat{\gamma _{\theta \theta }} \hat{\chi }'}{\hat{\alpha }^2 \hat{\chi }^2 \gamma _{\theta \theta }}+\frac{\hat{\beta ^r}^4 \hat{\gamma _{rr}}'}{2 \hat{\alpha }^2 \hat{\chi }}-\frac{{\beta^r} \hat{\beta ^r}^3 \hat{\gamma _{rr}}'}{\hat{\alpha }^2 \hat{\chi }}+\frac{{\beta^r}^2 \hat{\beta ^r}^2 \hat{\gamma _{rr}}'}{2 \hat{\alpha }^2 \hat{\chi }}+\frac{{\beta^r}^2 \hat{\gamma _{rr}}'}{2 \hat{\gamma _{rr}}}-\frac{\hat{\beta ^r}^2 \hat{\gamma _{rr}}'}{2 \hat{\gamma _{rr}}}-\frac{\alpha ^2 \chi  \hat{\beta ^r}^2 \hat{\gamma _{rr}}'}{2 \hat{\alpha }^2 \hat{\chi } \gamma _{rr}}
\nonumber \\* \fl &
-\frac{\alpha ^2 \chi  \hat{\gamma _{rr}}'}{2 \hat{\gamma _{rr}} \gamma _{rr}}-\frac{\alpha ^2 \chi  \hat{\beta ^r}^2 \hat{\gamma _{\theta \theta }}'}{\hat{\alpha }^2 \hat{\chi } \gamma _{\theta \theta }}+\frac{\alpha ^2 \chi  \hat{\gamma _{\theta \theta }}'}{\hat{\gamma _{rr}} \gamma _{\theta \theta }}+\frac{{\beta^r}^2 \gamma _{rr}'}{2 \gamma _{rr}}+\frac{\alpha ^2 \chi  \gamma _{rr}'}{2 \gamma _{rr}^2}+\frac{{\beta^r}^2 \gamma _{\theta \theta }'}{\gamma _{\theta \theta }}-\frac{\alpha ^2 \chi  \gamma _{\theta \theta }'}{\gamma _{rr} \gamma _{\theta \theta }}-\frac{2 \alpha ^2 \chi }{r \gamma _{rr}}
\nonumber \\* \fl &
-\frac{{\beta^r} \dot{\gamma _{rr}}}{2 \gamma _{rr}}-\frac{{\beta^r} \dot{\gamma _{\theta \theta }}}{\gamma _{\theta \theta }}-\frac{2 \alpha ^2 \chi  \hat{\beta ^r}^2 \hat{\gamma _{\theta \theta }}}{r \hat{\alpha }^2 \hat{\chi } \gamma _{\theta \theta }}+\frac{2 \alpha ^2 \chi  \hat{\gamma _{\theta \theta }}}{r \hat{\gamma _{rr}} \gamma _{\theta \theta }}
 .  \label{eg:harmdershiftbp}
\end{eqnarray}
\end{subequations}
\endgroup
}
The values of the background metric components $\hat\alpha$, $\hat{\beta^r}$, $\hat\chi$, $\hat{\gamma_{rr}}$ and $\hat{\gamma_{\theta\theta}}$ are given by \eref{e:hatvals}.
In the following simplified version of the previous equations, $\hat{\gamma_{rr}}$ and $\hat{\gamma_{\theta\theta}}$ are set to unity, the evolution equation $\dot\alpha$ is substituted in $\dot\beta^r$ and $\Lambda^r$ (determined by setting the $Z_r$ constraint, (C.2c) in \cite{\pap}, to zero) is used to eliminate the remaining $\gamma_{rr}'$ and $\gamma_{\theta\theta}'$ terms in $\dot\beta^r$ to make the system hyperbolic:
{\small
\begin{subequations}\label{eg:harmtwobpsimpl}
\begin{eqnarray}
\fl \dot\alpha =& {\beta^r} \alpha '-\alpha  {\beta^r}'-\frac{\alpha  {\beta^r} \gamma _{\theta \theta }'}{\gamma _{\theta \theta }}+\frac{3 \alpha  {\beta^r} \chi '}{2 \chi }+\frac{4 \alpha  {\beta^r} \Omega '}{\Omega }-\frac{2 \alpha  {\beta^r} \hat{\alpha }'}{\hat{\alpha }}+\frac{\alpha  \dot{\gamma _{\theta \theta }}}{\gamma _{\theta \theta }}-\frac{3 \alpha  \dot{\chi }}{2 \chi }-\frac{\alpha ^3 \tilde F^t}{\Omega ^2}
\nonumber \\ \fl &
-\frac{2 \alpha ^3 \chi  \hat{\beta ^r} \Omega '}{\hat{\alpha }^2 \hat{\chi } \Omega  \gamma _{\theta \theta }}-\frac{\alpha ^3 \chi  \hat{\beta ^r} \hat{\chi }'}{\hat{\alpha }^2 \hat{\chi }^2 \gamma _{\theta \theta }}+\frac{2 \alpha ^3 \chi  \hat{\beta ^r}}{\hat{\alpha }^2 r \hat{\chi } \gamma _{\theta \theta }}+\frac{\alpha  {\beta^r}^2 \hat{\beta ^r} \Omega '}{\hat{\alpha }^2 \hat{\chi } \Omega }-\frac{\alpha  {\beta^r}^2 \hat{\beta ^r}'}{\hat{\alpha }^2 \hat{\chi }}+\frac{\alpha  {\beta^r}^2 \hat{\beta ^r} \hat{\chi }'}{2 \hat{\alpha }^2 \hat{\chi }^2}-\frac{2 \alpha  {\beta^r} \hat{\beta ^r}^2 \Omega '}{\hat{\alpha }^2 \hat{\chi } \Omega }
\nonumber \\ \fl &
+\frac{2 \alpha  {\beta^r} \hat{\beta ^r} \hat{\beta ^r}'}{\hat{\alpha }^2 \hat{\chi }}-\frac{\alpha  {\beta^r} \hat{\beta ^r}^2 \hat{\chi }'}{\hat{\alpha }^2 \hat{\chi }^2}+\frac{\alpha  \hat{\beta ^r}^3 \Omega '}{\hat{\alpha }^2 \hat{\chi } \Omega }-\frac{\alpha  \hat{\beta ^r}^2 \hat{\beta ^r}'}{\hat{\alpha }^2 \hat{\chi }}+\frac{\alpha  \hat{\beta ^r}^3 \hat{\chi }'}{2 \hat{\alpha }^2 \hat{\chi }^2}-\frac{\alpha  \hat{\beta ^r} \Omega '}{\Omega }+\frac{\alpha  \hat{\alpha }' \hat{\beta ^r}}{\hat{\alpha }}-\frac{2 \alpha  {\beta^r}}{r}
\nonumber \\ \fl &
-\frac{\alpha ^3 \chi  \hat{\beta ^r} \Omega '}{\hat{\alpha }^2 \hat{\chi } \Omega  \gamma _{rr}}+\frac{\alpha ^3 \chi  \hat{\beta ^r}'}{\hat{\alpha }^2 \hat{\chi } \gamma _{rr}}-\frac{\alpha ^3 \chi  \hat{\beta ^r} \hat{\chi }'}{2 \hat{\alpha }^2 \hat{\chi }^2 \gamma _{rr}}-\frac{\alpha  {\beta^r} \gamma _{rr}'}{2 \gamma _{rr}}+\frac{\alpha  \dot{\gamma _{rr}}}{2 \gamma _{rr}}
 , \label{eg:harmderlapsebpsimpl} \\
\fl \dot \beta^r =&  -\frac{\alpha ^2 \tilde F^r}{\Omega ^2}-\frac{\alpha ^2 {\beta^r} \tilde F^t}{\Omega ^2}+\alpha ^2 \Lambda ^r \chi -\frac{\alpha  \chi  \alpha '}{\gamma _{rr}}+{\beta^r} {\beta^r}'+\frac{\alpha ^2 \chi '}{2 \gamma _{rr}}+\frac{2 \hat{\beta ^r}^2 \Omega '}{\Omega }-\frac{3 {\beta^r} \hat{\beta ^r} \Omega '}{\Omega }-\frac{\hat{\alpha }^2 \hat{\chi } \Omega '}{\Omega }
\nonumber \\ \fl &
+\frac{{\beta^r}^2 \Omega '}{\Omega }-\frac{\hat{\beta ^r}^4 \Omega '}{\Omega  \hat{\alpha }^2 \hat{\chi }}+\frac{3 {\beta^r} \hat{\beta ^r}^3 \Omega '}{\Omega  \hat{\alpha }^2 \hat{\chi }}-\frac{3 {\beta^r}^2 \hat{\beta ^r}^2 \Omega '}{\Omega  \hat{\alpha }^2 \hat{\chi }}+\frac{{\beta^r}^3 \hat{\beta ^r} \Omega '}{\Omega  \hat{\alpha }^2 \hat{\chi }}+\frac{3 \alpha ^2 \chi  \Omega '}{\Omega  \gamma _{rr}}+\frac{\alpha ^2 \chi  \hat{\beta ^r}^2 \Omega '}{\Omega  \hat{\alpha }^2 \hat{\chi } \gamma _{rr}}
\nonumber \\ \fl &
-\frac{\alpha ^2 {\beta^r} \chi  \hat{\beta ^r} \Omega '}{\Omega  \hat{\alpha }^2 \hat{\chi } \gamma _{rr}}-\frac{2 \alpha ^2 \chi  \Omega '}{\Omega  \gamma _{\theta \theta }}+\frac{2 \alpha ^2 \chi  \hat{\beta ^r}^2 \Omega '}{\Omega  \hat{\alpha }^2 \hat{\chi } \gamma _{\theta \theta }}-\frac{2 \alpha ^2 {\beta^r} \chi  \hat{\beta ^r} \Omega '}{\Omega  \hat{\alpha }^2 \hat{\chi } \gamma _{\theta \theta }}-\frac{\hat{\beta ^r}^2 \hat{\alpha }'}{\hat{\alpha }}+\frac{3 {\beta^r} \hat{\beta ^r} \hat{\alpha }'}{\hat{\alpha }}+\hat{\alpha } \hat{\chi } \hat{\alpha }'
\nonumber \\ \fl &
-\frac{2 {\beta^r}^2 \hat{\alpha }'}{\hat{\alpha }}-\hat{\beta ^r} \hat{\beta ^r}'+\frac{\hat{\beta ^r}^3 \hat{\beta ^r}'}{\hat{\alpha }^2 \hat{\chi }}-\frac{3 {\beta^r} \hat{\beta ^r}^2 \hat{\beta ^r}'}{\hat{\alpha }^2 \hat{\chi }}+\frac{3 {\beta^r}^2 \hat{\beta ^r} \hat{\beta ^r}'}{\hat{\alpha }^2 \hat{\chi }}-\frac{{\beta^r}^3 \hat{\beta ^r}'}{\hat{\alpha }^2 \hat{\chi }}-\frac{\alpha ^2 \chi  \hat{\beta ^r} \hat{\beta ^r}'}{\hat{\alpha }^2 \hat{\chi } \gamma _{rr}}+\frac{\alpha ^2 {\beta^r} \chi  \hat{\beta ^r}'}{\hat{\alpha }^2 \hat{\chi } \gamma _{rr}}
\nonumber \\ \fl &
-\frac{{\beta^r}^2 \hat{\chi }'}{2 \hat{\chi }}+\frac{\hat{\beta ^r}^2 \hat{\chi }'}{2 \hat{\chi }}+\frac{\alpha ^2 \chi  \hat{\chi }'}{2 \hat{\chi } \gamma _{rr}}+\frac{\alpha ^2 \chi  \hat{\beta ^r}^2 \hat{\chi }'}{2 \hat{\alpha }^2 \hat{\chi }^2 \gamma _{rr}}-\frac{\alpha ^2 {\beta^r} \chi  \hat{\beta ^r} \hat{\chi }'}{2 \hat{\alpha }^2 \hat{\chi }^2 \gamma _{rr}}-\frac{\alpha ^2 \chi  \hat{\chi }'}{\hat{\chi } \gamma _{\theta \theta }}+\frac{\alpha ^2 \chi  \hat{\beta ^r}^2 \hat{\chi }'}{\hat{\alpha }^2 \hat{\chi }^2 \gamma _{\theta \theta }}-\frac{\alpha ^2 {\beta^r} \chi  \hat{\beta ^r} \hat{\chi }'}{\hat{\alpha }^2 \hat{\chi }^2 \gamma _{\theta \theta }}
\nonumber \\ \fl &
-\frac{\hat{\beta ^r}^4 \hat{\chi }'}{2 \hat{\alpha }^2 \hat{\chi }^2}+\frac{3 {\beta^r} \hat{\beta ^r}^3 \hat{\chi }'}{2 \hat{\alpha }^2 \hat{\chi }^2}-\frac{3 {\beta^r}^2 \hat{\beta ^r}^2 \hat{\chi }'}{2 \hat{\alpha }^2 \hat{\chi }^2}+\frac{{\beta^r}^3 \hat{\beta ^r} \hat{\chi }'}{2 \hat{\alpha }^2 \hat{\chi }^2}-\frac{2 \alpha ^2 \chi  \hat{\beta ^r}^2}{r \hat{\alpha }^2 \hat{\chi } \gamma _{\theta \theta }}+\frac{2 \alpha ^2 {\beta^r} \chi  \hat{\beta ^r}}{r \hat{\alpha }^2 \hat{\chi } \gamma _{\theta \theta }}
 .  \label{eg:harmdershiftbpsimpl}
\end{eqnarray}
\end{subequations}
}%
In the code we set the free specifiable source functions to $\tilde F^r=0$ and $\tilde F^t = \cpbg\Omega(\alpha-\hat\alpha)$. The equation of motion for $\dot\chi$ ($\gamma_{\theta\theta}$ is eliminated in terms of $\gamma_{rr}$, which makes the $\dot\gamma_{rr}$ and $\dot\gamma_{\theta\theta}$ terms cancel) is to be substituted above.

\section{Previous slicing conditions}\label{a:oldlapse}

Slicing condition \eref{ae:masterphys} makes it easy to implement the harmonic, 1+log and cK conditions (or a matching between them), by providing source functions and damping terms that are valid for any of them.
Here we present some of the previous equations of motion for the lapse that were used in our implementation before the simple equation \eref{ae:masterphys} was found.
In general, $\xi=1,2$ is a good choice for the $\charm$, $\cuplog$ and $\cnK$ parameters.

\subsection{Defined in the physical domain:}

The harmonic slicing takes the following form:
\begin{eqnarray}\label{ae:harmp}
\fl \dot{\alpha} =& {\beta^r} \alpha '-\backbetar \backalpha'-\frac{\alpha ^2 \Delta\tilde K}{\Omega }+\frac{\charm-\Kc}{\Omega} \left(\backalpha^2-\alpha ^2\right)+\left(\frac{\backalpha  \backbetar \Omega'}{\Omega }-\frac{{\alpha} {\beta^r} \Omega '}{\Omega }\right).
\end{eqnarray}
The $\Delta\tilde K$ term in the 1+log condition that follows has been divided by $\Omega$ in order to prevent the lapse gauge speeds from vanishing at $\scri^+$ (this is an effect of the transformation from the physical to the conformal domain):
\begin{eqnarray}\label{ae:1plogp}
\fl    \dot{\alpha} =& {\beta^r} \alpha '-\backbetar\backalpha'-\frac{n_{1+log} \alpha  \Delta\tilde K}{\Omega }+\frac{\cuplog n_{1+log}}{\Omega }\left(\backalpha-\alpha\right)+ \left(\frac{\backalpha \backbetar \Omega '}{\Omega }-\frac{\alpha{\beta^r} \Omega '}{\Omega }\right) .
\end{eqnarray}
In the cK case, it is by $\Omega^2$ that the $\Delta \tilde K$ terms needs to be divided:
\begin{eqnarray}\label{ae:constp}
\fl \dot{\alpha} =& {\beta^r} \alpha '-\backbetar \backalpha'-\frac{\nconst \Delta\tilde K}{\Omega }+\frac{\cnK }{ \Omega }\left(\backalpha-\alpha\right)+\left(\frac{\backalpha  \backbetar \Omega '}{\Omega }-\frac{\alpha {\beta^r} \Omega '}{\Omega }\right) .
\end{eqnarray}

\subsection{Defined in the conformal domain:}\label{as:lapseconf}

The difference with the previous set of equations is that the last damping term has a different coefficient (3 instead of 1).
The harmonic gauge is implemented as:
\begin{eqnarray}\label{ae:harm}
\fl \dot{\alpha} =& {\beta^r} \alpha '-\backbetar \backalpha'-\frac{\alpha ^2 \Delta\tilde K}{\Omega }+\charm \left(\backalpha^2-\alpha ^2\right)+\left(\frac{3 \backalpha  \backbetar \Omega'}{\Omega }-\frac{3 {\alpha} {\beta^r} \Omega '}{\Omega }\right).
\end{eqnarray}
Again, the damping is controlled by the parameter $\charm$.
In the 1+log condition that follows, the equivalent damping term formally diverges at $\scri^+$:
\begin{eqnarray}\label{ae:1plog}
\fl    \dot{\alpha} =& {\beta^r} \alpha '-\backbetar\backalpha'-\frac{n_{1+log} \alpha  \Delta\tilde K}{\Omega }+\frac{\cuplog \Kc}{3 \Omega }\left(\backalpha-\alpha\right)+n_{1+log} \left(\frac{3 \backbetar \Omega '}{\Omega }-\frac{3 {\beta^r} \Omega '}{\Omega }\right) .
\end{eqnarray}
The same is valid for the cK condition, whose  difference with the 1+log one is the coefficient in front of $\Delta\tilde K$:
\begin{eqnarray}\label{ae:const}
\fl \dot{\alpha} =& {\beta^r} \alpha '-\backbetar \backalpha'-\frac{\nconst \Delta\tilde K}{\Omega }+\frac{\cnK \Kc }{3 \Omega }\left(\backalpha-\alpha\right)+\nconst\left(\frac{3 \backbetar \Omega '}{\backalpha \Omega }-\frac{3 {\beta^r} \Omega '}{\alpha  \Omega }\right) .
\end{eqnarray}

\section*{References}
\bibliographystyle{unsrt}
\bibliography{hypcomp}

\begin{thebibliography}{10}

\bibitem{Barack:1998bv}
Leor Barack.
\newblock {Late time dynamics of scalar perturbations outside black holes. 1. A
  Shell toy model}.
\newblock {\em Phys.Rev.}, D59:044016, 1999.

\bibitem{Leaver1986}
Edward~W. Leaver.
\newblock {Solutions to a generalized spheroidal wave equation}.
\newblock {\em J.Math.Phys.}, 27:1238, 1986.

\bibitem{PhysRevD.34.384}
Edward~W. Leaver.
\newblock {Spectral decomposition of the perturbation response of the
  Schwarzschild geometry}.
\newblock {\em Phys. Rev. D}, 34:384--408, Jul 1986.

\bibitem{PhysRevLett.10.66}
Roger Penrose.
\newblock Asymptotic properties of fields and space-times.
\newblock {\em Phys. Rev. Lett.}, 10:66--68, Jan 1963.

\bibitem{Penrose:1965am}
Roger Penrose.
\newblock {Zero rest mass fields including gravitation: Asymptotic behavior}.
\newblock {\em Proc.Roy.Soc.Lond.}, A284:159, 1965.

\bibitem{friedrich1983}
Helmut Friedrich.
\newblock Cauchy problems for the conformal vacuum field equations in general
  relativity.
\newblock {\em Comm. Math. Phys.}, 91(4):445--472, 1983.

\bibitem{lrr-2004-1}
J\"org Frauendiener.
\newblock Conformal infinity.
\newblock {\em Living Reviews in Relativity}, 7(1), 2004.

\bibitem{Friedrich:2003fq}
Helmut Friedrich.
\newblock {\em Conformal Einstein Evolution}, pages 1--50.
\newblock Springer Berlin Heidelberg, Berlin, Heidelberg, 2002.

\bibitem{Husa:2002zc}
Sascha Husa.
\newblock {\em Numerical Relativity with the Conformal Field Equations}, volume
  617, pages 159--192.
\newblock Springer Berlin Heidelberg, Berlin, Heidelberg, 2003.

\bibitem{Husa:2005ns}
Sascha Husa, Carsten Schneemann, Tilman Vogel, and An{\i}l Zengino\u{g}lu.
\newblock {Hyperboloidal data and evolution}.
\newblock {\em AIP Conf.Proc.}, 841:306--313, 2006.

\bibitem{Frauendiener:2004bj}
J\"org Frauendiener and Tilman Vogel.
\newblock {Algebraic stability analysis of constraint propagation}.
\newblock {\em Class.Quant.Grav.}, 22:1769--1793, 2005.

\bibitem{Abbott:2016blz}
B.~P. Abbott et~al.
\newblock {Observation of Gravitational Waves from a Binary Black Hole Merger}.
\newblock {\em Phys. Rev. Lett.}, 116(6):061102, 2016.

\bibitem{Abbott:2016nmj}
B.~P. Abbott et~al.
\newblock {GW151226: Observation of Gravitational Waves from a 22-Solar-Mass
  Binary Black Hole Coalescence}.
\newblock {\em Phys. Rev. Lett.}, 116(24):241103, 2016.

\bibitem{TheLIGOScientific:2016wfe}
B.~P. Abbott et~al.
\newblock {Properties of the Binary Black Hole Merger GW150914}.
\newblock {\em Phys. Rev. Lett.}, 116(24):241102, 2016.

\bibitem{TheLIGOScientific:2016pea}
B.~P. Abbott et~al.
\newblock {Binary Black Hole Mergers in the first Advanced LIGO Observing Run}.
\newblock {\em Phys. Rev.}, X6(4):041015, 2016.

\bibitem{Boyle:2009vi}
Michael Boyle and Abdul~H. Mroue.
\newblock {Extrapolating gravitational-wave data from numerical simulations}.
\newblock {\em Phys.Rev.}, D80:124045, 2009.

\bibitem{Reisswig:2009us}
C.~Reisswig, N.T. Bishop, D.~Pollney, and B.~Szilagyi.
\newblock {Unambiguous determination of gravitational waveforms from binary
  black hole mergers}.
\newblock {\em Phys.Rev.Lett.}, 103:221101, 2009.

\bibitem{Pretorius:2005gq}
Frans Pretorius.
\newblock {Evolution of binary black hole spacetimes}.
\newblock {\em Phys. Rev. Lett.}, 95:121101, 2005.

\bibitem{Campanelli:2005dd}
Manuela Campanelli, C.O. Lousto, P.~Marronetti, and Y.~Zlochower.
\newblock {Accurate evolutions of orbiting black-hole binaries without
  excision}.
\newblock {\em Phys.Rev.Lett.}, 96:111101, 2006.

\bibitem{Baker:2005vv}
John~G. Baker, Joan Centrella, Dae-Il Choi, Michael Koppitz, and James van
  Meter.
\newblock {Gravitational wave extraction from an inspiraling configuration of
  merging black holes}.
\newblock {\em Phys.Rev.Lett.}, 96:111102, 2006.

\bibitem{PhysRevD.52.5428}
Masaru Shibata and Takashi Nakamura.
\newblock Evolution of three-dimensional gravitational waves: Harmonic slicing
  case.
\newblock {\em Phys. Rev. D}, 52:5428--5444, Nov 1995.

\bibitem{Baumgarte:1998te}
Thomas~W. Baumgarte and Stuart~L. Shapiro.
\newblock {On the numerical integration of Einstein's field equations}.
\newblock {\em Phys.Rev.}, D59:024007, 1999.

\bibitem{bona-2003-67}
C.~Bona, T.~Ledvinka, C.~Palenzuela, and M.~Zacek.
\newblock {General-covariant evolution formalism for Numerical Relativity}.
\newblock {\em {Phys. Rev.}}, D67:104005, 2003.

\bibitem{Hannam:2006vv}
Mark Hannam, Sascha Husa, Denis Pollney, Bernd Br\"{u}gmann, and Niall
  O'Murchadha.
\newblock {Geometry and regularity of moving punctures}.
\newblock {\em Phys.Rev.Lett.}, 99:241102, 2007.

\bibitem{Hannam:2008sg}
Mark Hannam, Sascha Husa, Frank Ohme, Bernd Br\"{u}gmann, and Niall
  O'Murchadha.
\newblock {Wormholes and trumpets: The Schwarzschild spacetime for the
  moving-puncture generation}.
\newblock {\em Phys.Rev.}, D78:064020, 2008.

\bibitem{Choptuik:1992jv}
Matthew~W. Choptuik.
\newblock {Universality and scaling in gravitational collapse of a massless
  scalar field}.
\newblock {\em Phys.Rev.Lett.}, 70:9--12, 1993.

\bibitem{Tamburino:1966zz}
Louis~A. Tamburino and Jeffrey~H. Winicour.
\newblock {Gravitational Fields in Finite and Conformal Bondi Frames}.
\newblock {\em Phys.Rev.}, 150:1039--1053, 1966.

\bibitem{9780511564048}
Roger Penrose and Wolfgang Rindler.
\newblock {\em Spinors and Space-Time}, volume 1, 2.
\newblock Cambridge University Press, 1984, 1986.
\newblock Cambridge Books Online.

\bibitem{stewart1997advanced}
John Stewart.
\newblock {\em Advanced General Relativity}.
\newblock Mir, 1997.

\bibitem{Andersson:springer}
Lars Andersson.
\newblock Construction of hyperboloidal initial data.
\newblock In J\"org Frauendiener and Helmut Friedrich, editors, {\em The
  Conformal Structure of Space-Time}, volume 604 of {\em Lecture Notes in
  Physics}, pages 183--194. Springer Berlin Heidelberg, 2002.

\bibitem{Rinne:2009qx}
Oliver Rinne.
\newblock {An Axisymmetric evolution code for the Einstein equations on
  hyperboloidal slices}.
\newblock {\em Class.Quant.Grav.}, 27:035014, 2010.

\bibitem{Rinne:2013qc}
Oliver Rinne and Vincent Moncrief.
\newblock {Hyperboloidal Einstein-matter evolution and tails for scalar and
  Yang-Mills fields}.
\newblock {\em Class.Quant.Grav.}, 30:095009, 2013.

\bibitem{Bardeen:2011ip}
James~M. Bardeen, Olivier Sarbach, and Luisa~T. Buchman.
\newblock {Tetrad formalism for numerical relativity on conformally
  compactified constant mean curvature hypersurfaces}.
\newblock {\em Phys.Rev.}, D83:104045, 2011.

\bibitem{Morales:2016rgt}
Manuel~D. Morales and Olivier Sarbach.
\newblock {Evolution of scalar fields surrounding black holes on compactified
  constant mean curvature hypersurfaces}.
\newblock {\em Phys. Rev.}, D95(4):044001, 2017.

\bibitem{Zenginoglu:2007jw}
An{\i}l Zengino\u{g}lu.
\newblock {Hyperboloidal foliations and scri-fixing}.
\newblock {\em Class.Quant.Grav.}, 25:145002, 2008.

\bibitem{Zenginoglu:2008pw}
An{\i}l Zengino\u{g}lu.
\newblock {Hyperboloidal evolution with the Einstein equations}.
\newblock {\em Class.Quant.Grav.}, 25:195025, 2008.

\bibitem{Zenginoglu:2008wc}
An{\i}l Zengino\u{g}lu.
\newblock {A Hyperboloidal study of tail decay rates for scalar and Yang-Mills
  fields}.
\newblock {\em Class.Quant.Grav.}, 25:175013, 2008.

\bibitem{Zenginoglu:2007it}
An{\i}l Zengino\u{g}lu.
\newblock {\em {A conformal approach to numerical calculations of
  asymptotically flat spacetimes}}.
\newblock PhD thesis,
  Max Planck Institute for Gravitational Physics (AEI) and University of
  Potsdam, Institute of Physics and Astronomy, 2007.

\bibitem{Vano-Vinuales:2014koa}
Alex Va{\~n}{\'o}-Vi{\~n}uales, Sascha Husa, and David Hilditch.
\newblock {Spherical symmetry as a test case for unconstrained hyperboloidal
  evolution}.
\newblock {\em Class. Quant. Grav.}, 32(17):175010, 2015.

\bibitem{bona}
Carles Bona, Carlos Palenzuela-Luque, and Carles Bona-Casas.
\newblock {\em {Elements of Numerical Relativity and Relativistic
  Hydrodynamics}}.
\newblock Springer, second edition, 2009.

\bibitem{Bona:1994dr}
Carles Bona, Joan Mass\'o, Edward Seidel, and Joan Stela.
\newblock {A New formalism for numerical relativity}.
\newblock {\em Phys.Rev.Lett.}, 75:600--603, 1995.

\bibitem{Alcubierre:2002kk}
Miguel Alcubierre, Bernd Br\"{u}gmann, Peter Diener, Michael Koppitz, Denis
  Pollney, et~al.
\newblock {Gauge conditions for long term numerical black hole evolutions
  without excision}.
\newblock {\em Phys.Rev.}, D67:084023, 2003.

\bibitem{Friedrich:2000qv}
Helmut Friedrich and Alan~D. Rendall.
\newblock {\em {The Cauchy Problem for the Einstein Equations}}, volume 540,
  pages 127--224.
\newblock Springer Berlin Heidelberg, Berlin, Heidelberg, 2000.

\bibitem{Ohme:2009gn}
Frank Ohme, Mark Hannam, Sascha Husa, and Niall O'Murchadha.
\newblock {Stationary hyperboloidal slicings with evolved gauge conditions}.
\newblock {\em Class.Quant.Grav.}, 26:175014, 2009.

\bibitem{bhpaper}
Alex Va{\~n}{\'o}-Vi{\~n}uales and Sascha Husa.
\newblock {Spherical symmetry as a test case for unconstrained hyperboloidal
  evolution III: black holes}.
\newblock 2017.
\newblock in preparation.

\bibitem{Buchman:2009ew}
Luisa~T. Buchman, Harald~P. Pfeiffer, and James~M. Bardeen.
\newblock {Black hole initial data on hyperboloidal slices}.
\newblock {\em Phys.Rev.}, D80:084024, 2009.

\bibitem{Hilditch:2015qea}
David Hilditch.
\newblock {Dual Foliation Formulations of General Relativity}.
\newblock 2015.

\bibitem{Hilditch:2016xzh}
David Hilditch, Enno Harms, Marcus Bugner, Hannes Rueter, and Bernd Bruegmann.
\newblock {The evolution of hyperboloidal data with the dual foliation
  formalism: Mathematical analysis and wave equation tests}.
\newblock 2016.

\bibitem{Schneemann}
Carsten Schneemann.
\newblock {Numerische Berechnung von hyperboloidalen Anfangsdaten f\"ur die
  Einstein-Gleichungen}.
\newblock Master's thesis, 2006.

\bibitem{Zenginoglu:2006rj}
Anil Zenginoglu and Sascha Husa.
\newblock {Hyperboloidal foliations with scri-fixing in spherical symmetry}.
\newblock In {\em {Recent developments in theoretical and experimental general
  relativity, gravitation and relativistic field theories. Proceedings, 11th
  Marcel Grossmann Meeting, MG11, Berlin, Germany, July 23-29, 2006. Pt. A-C}},
  pages 1624--1626, 2006.

\bibitem{Zenginoglu:2008uc}
An{\i}l Zengino\u{g}lu, Dario Nunez, and Sascha Husa.
\newblock {Gravitational perturbations of Schwarzschild spacetime at null
  infinity and the hyperboloidal initial value problem}.
\newblock {\em Class.Quant.Grav.}, 26:035009, 2009.

\bibitem{Frauendiener:1997ze}
J\"org Frauendiener.
\newblock {Numerical treatment of the hyperboloidal initial value problem for
  the vacuum Einstein equations. 2. The Evolution equations}.
\newblock {\em Phys.Rev.}, D58:064003, 1998.

\bibitem{Wald}
Robert~M. Wald.
\newblock {\em {General Relativity}}.
\newblock The University of Chicago Press, 1984.

\bibitem{Vano-Vinuales:2015lhj}
Alex Va{\~n}{\'o}-Vi{\~n}uales.
\newblock {\em {Free evolution of the hyperboloidal initial value problem in
  spherical symmetry}}.
\newblock PhD thesis, U. Illes Balears, Palma, 2015.

\bibitem{newman1962approach}
Ezra Newman and Roger Penrose.
\newblock An approach to gravitational radiation by a method of spin
  coefficients.
\newblock {\em Journal of Mathematical Physics}, 3(3):566--578, 1962.

\bibitem{gustafsson1995time}
B.~Gustafsson, H.O. Kreiss, and J.~Oliger.
\newblock {\em Time dependent problems and difference methods}.
\newblock Pure and applied mathematics. Wiley, 1995.

\bibitem{Calabrese:2005ft}
Gioel Calabrese, Ian Hinder, and Sascha Husa.
\newblock {Numerical stability for finite difference approximations of
  Einstein's equations}.
\newblock {\em J.Comput.Phys.}, 218:607--634, 2006.

\bibitem{Husa:2007zz}
Sascha Husa.
\newblock {Numerical modeling of black holes as sources of gravitational waves
  in a nutshell}.
\newblock {\em Eur. Phys. J. ST}, 152:183--207, 2007.

\bibitem{Arbona:1999ym}
A.~Arbona, C.~Bona, J.~Mass\'o, and J.~Stela.
\newblock {Robust evolution system for numerical relativity}.
\newblock {\em Phys.Rev.}, D60:104014, 1999.

\bibitem{Alcubierre:2001vm}
Miguel Alcubierre, Bernd Br\"{u}gmann, Denis Pollney, Edward Seidel, and Ryoji
  Takahashi.
\newblock {Black hole excision for dynamic black holes}.
\newblock {\em Phys.Rev.}, D64:061501, 2001.

\bibitem{Vano-Vinuales:2014ada}
Alex Va{\~n}{\'o}-Vi{\~n}uales and Sascha Husa.
\newblock {Unconstrained hyperboloidal evolution of black holes in spherical
  symmetry with GBSSN and Z4c}.
\newblock {\em J.Phys.Conf.Ser.}, 600(1):012061, 2015.

\bibitem{Alcubierre:1996su}
Miguel Alcubierre.
\newblock {The Appearance of coordinate shocks in hyperbolic formalisms of
  general relativity}.
\newblock {\em Phys. Rev.}, D55:5981--5991, 1997.

\bibitem{Alcubierre:2002iq}
Miguel Alcubierre.
\newblock {Hyperbolic slicings of space-time: Singularity avoidance and gauge
  shocks}.
\newblock {\em Class. Quant. Grav.}, 20:607--624, 2003.

\bibitem{Reimann:2004wp}
Bernd Reimann, Miguel Alcubierre, Jose~A. Gonzalez, and Dario Nunez.
\newblock {Constraint and gauge shocks in one-dimensional numerical
  relativity}.
\newblock {\em Phys. Rev.}, D71:064021, 2005.

\bibitem{Alcubierre:2005gw}
Miguel Alcubierre.
\newblock {Are gauge shocks really shocks?}
\newblock {\em Class. Quant. Grav.}, 22:4071--4082, 2005.

\bibitem{Garfinkle:2007yt}
David Garfinkle, Carsten Gundlach, and David Hilditch.
\newblock {Comments on Bona-Masso type slicing conditions in long-term black
  hole evolutions}.
\newblock {\em Class. Quant. Grav.}, 25:075007, 2008.

\bibitem{1993PhDT.........3B}
D.~H. {Bernstein}.
\newblock {\em {Physics, astronomy, and astrophysics: A numerical study of the
  black hole plus Brill wave spacetime}}.
\newblock PhD thesis, Illinois Univ., Urbana-Champaign., 1993.

\bibitem{Anninos:1995am}
Peter Anninos, Karen Camarda, Joan Masso, Edward Seidel, Wai-Mo Suen, et~al.
\newblock {Three-dimensional numerical relativity: The Evolution of black
  holes}.
\newblock {\em Phys.Rev.}, D52:2059--2082, 1995.

\bibitem{Brown:2009dd}
J.~David Brown.
\newblock {Covariant formulations of BSSN and the standard gauge}.
\newblock {\em Phys. Rev.}, D79:104029, 2009.

\bibitem{vanMeter:2006vi}
James~R. van Meter, John~G. Baker, Michael Koppitz, and Dae-Il Choi.
\newblock {How to move a black hole without excision: Gauge conditions for the
  numerical evolution of a moving puncture}.
\newblock {\em Phys.Rev.}, D73:124011, 2006.

\bibitem{Gundlach:2006tw}
Carsten Gundlach and Jose~M. Martin-Garcia.
\newblock {Well-posedness of formulations of the Einstein equations with
  dynamical lapse and shift conditions}.
\newblock {\em Phys.Rev.}, D74:024016, 2006.

\bibitem{Hilditch:2013ila}
David Hilditch and Ronny Richter.
\newblock {Hyperbolicity of Physical Theories with Application to General
  Relativity}.
\newblock {\em Phys. Rev.}, D94(4):044028, 2016.

\bibitem{Alcubierre:2005gh}
Miguel Alcubierre, Alejandro Corichi, Jose~A. Gonzalez, Dario Nunez, Bernd
  Reimann, et~al.
\newblock {Generalized harmonic spatial coordinates and hyperbolic shift
  conditions}.
\newblock {\em Phys.Rev.}, D72:124018, 2005.

\bibitem{Husa:2007hp}
Sascha Husa, Jose~A. Gonzalez, Mark Hannam, Bernd Br\"{u}gmann, and Ulrich
  Sperhake.
\newblock {Reducing phase error in long numerical binary black hole evolutions
  with sixth order finite differencing}.
\newblock {\em Class.Quant.Grav.}, 25:105006, 2008.

\bibitem{kreiss1973methods}
H.O. Kreiss and J.~Oliger.
\newblock {\em Methods for the approximate solution of time dependent
  problems}.
\newblock GARP publications series No. 10. International Council of Scientific
  Unions, World Meteorological Organization, 1973.

\bibitem{Babiuc:2007vr}
M.C. Babiuc, S.~Husa, D.~Alic, I.~Hinder, C.~Lechner, et~al.
\newblock {Implementation of standard testbeds for numerical relativity}.
\newblock {\em Class.Quant.Grav.}, 25:125012, 2008.

\bibitem{Brandt:1997tf}
Steven Brandt and Bernd Br\"{u}gmann.
\newblock {A Simple construction of initial data for multiple black holes}.
\newblock {\em Phys.Rev.Lett.}, 78:3606--3609, 1997.

\bibitem{Vano-Vinuales:2016mbo}
Alex Vañó-Viñuales and Sascha Husa.
\newblock {Free hyperboloidal evolution in spherical symmetry}.
\newblock 2016.

\bibitem{Bruegmann:1997uc}
Bernd Br\"{u}gmann.
\newblock {Binary black hole mergers in 3-d numerical relativity}.
\newblock {\em Int.J.Mod.Phys.}, D8:85, 1999.

\bibitem{Bruegmann:2006at}
Bernd Br\"{u}gmann, Jose~A. Gonzalez, Mark Hannam, Sascha Husa, Ulrich
  Sperhake, et~al.
\newblock {Calibration of Moving Puncture Simulations}.
\newblock {\em Phys.Rev.}, D77:024027, 2008.

\bibitem{xAct}
José~M. Martín-García.
\newblock {{\em xAct:} Efficient tensor computer algebra for {\it
  Mathematica}}.
\newblock {\tt http://www.xact.es/}.

\bibitem{Brown:2007nt}
J.~David Brown.
\newblock {BSSN in Spherical Symmetry}.
\newblock {\em Class. Quant. Grav.}, 25:205004, 2008.

\bibitem{Alic:2011gg}
Daniela Alic, Carles Bona-Casas, Carles Bona, Luciano Rezzolla, and Carlos
  Palenzuela.
\newblock {Conformal and covariant formulation of the Z4 system with
  constraint-violation damping}.
\newblock {\em Phys.Rev.}, D85:064040, 2012.

\bibitem{Sanchis-Gual:2014nha}
Nicolas Sanchis-Gual, Pedro~J. Montero, Jose~A. Font, Ewald M\"uller, and
  Thomas~W. Baumgarte.
\newblock {Fully covariant and conformal formulation of the Z4 system in a
  reference-metric approach: comparison with the BSSN formulation in spherical
  symmetry}.
\newblock {\em Phys. Rev.}, D89(10):104033, 2014.

\bibitem{Bernuzzi:2009ex}
Sebastiano Bernuzzi and David Hilditch.
\newblock {Constraint violation in free evolution schemes: Comparing BSSNOK
  with a conformal decomposition of Z4}.
\newblock {\em Phys.Rev.}, D81:084003, 2010.

\bibitem{Weyhausen:2011cg}
Andreas Weyhausen, Sebastiano Bernuzzi, and David Hilditch.
\newblock {Constraint damping for the Z4c formulation of general relativity}.
\newblock {\em Phys.Rev.}, D85:024038, 2012.

\end{thebibliography}

\end{document}